%% file: root.tex
\documentclass[10pt,twocolumn,letterpaper]{article}

\usepackage[dvipsnames]{xcolor}
\usepackage[pdftex]{graphicx}
\usepackage{tikz}
\usetikzlibrary{calc}

\usepackage{cvpr}
\usepackage{times}

\usepackage[english]{babel}
\usepackage[english]{translator}

\usepackage[utf8]{inputenc}
\usepackage[T1]{fontenc} 

\input{./misc/preamble/preamble}
\input{./misc/notations/maths}
\input{./misc/notations/notations}

\input{./misc/notations/glossary}
\cvprfinalcopy %

\def\cvprPaperID{9392} %

\title{Blur Aware Calibration of Multi-Focus Plenoptic Camera}

\definecolor{somegray}{rgb}{0.5, 0.5, 0.5}
\newcommand{\darkgrayed}[1]{\textcolor{somegray}{#1}}
\makeatletter
\newcommand*\titleheader[1]{\gdef\@titleheader{#1}}
\AtBeginDocument{%
	\let\st@red@title\@title
	\def\@title{%
		\vskip-4em
		\bgroup\normalfont\large\centering\@titleheader\par\egroup
		\vskip1.5em\st@red@title}
}
\makeatother

\titleheader{\darkgrayed{This paper has been accepted for publication at the\\
		IEEE Conference on Computer Vision and Pattern Recognition (CVPR), Seattle, 2020.
		\copyright IEEE}}

\title{Blur Aware Calibration of Multi-Focus Plenoptic Camera}

\ifcvprfinal\fi
\begin{document}

\author{Mathieu Labussi\`{e}re$^1$, C\'{e}line Teuli\`{e}re$^1$, Fr\'{e}d\'{e}ric Bernardin$^2$, Omar Ait-Aider$^1$ \\ %
\noindent$^1$ Universit\'{e} Clermont Auvergne, CNRS, SIGMA Clermont,\\ Institut Pascal,
	F-63000 Clermont-Ferrand, France\\ 
\noindent$^2$ Cerema, \'{E}quipe-projet STI, 10 rue Bernard Palissy, F-63017 Clermont-Ferrand, France\\
	{\tt\small mathieu.labu@gmail.com, firstname.name@\{uca,cerema\}.fr}
}

\maketitle
\thispagestyle{empty}

\begin{abstract}
	This paper presents a novel calibration algorithm for \glspl{MFPC} using raw images only. %
	The design of such cameras is usually complex and relies on precise placement of optic elements. %
	Several calibration procedures have been proposed to retrieve the camera parameters but relying on simplified models, reconstructed images to extract features, or multiple calibrations when several types of micro-lens are used.
	Considering blur information, we propose a new \acrfull{BAP} feature. 
	It is first exploited in a pre-calibration step that retrieves initial camera parameters, %
	and secondly to express a new cost function for our single optimization process.
	The effectiveness of our calibration method is validated by quantitative and qualitative experiments.%
\end{abstract}
\glsresetall
\vspace*{-2mm}
\section{Introduction}

\begin{figure}[t]
	\renewcommand\fbox{\fcolorbox{white}{white}}
	\setlength{\fboxsep}{0pt}%
	\setlength{\fboxrule}{1.5pt}%
	\begin{center}
		\includegraphics[width=0.95\linewidth]{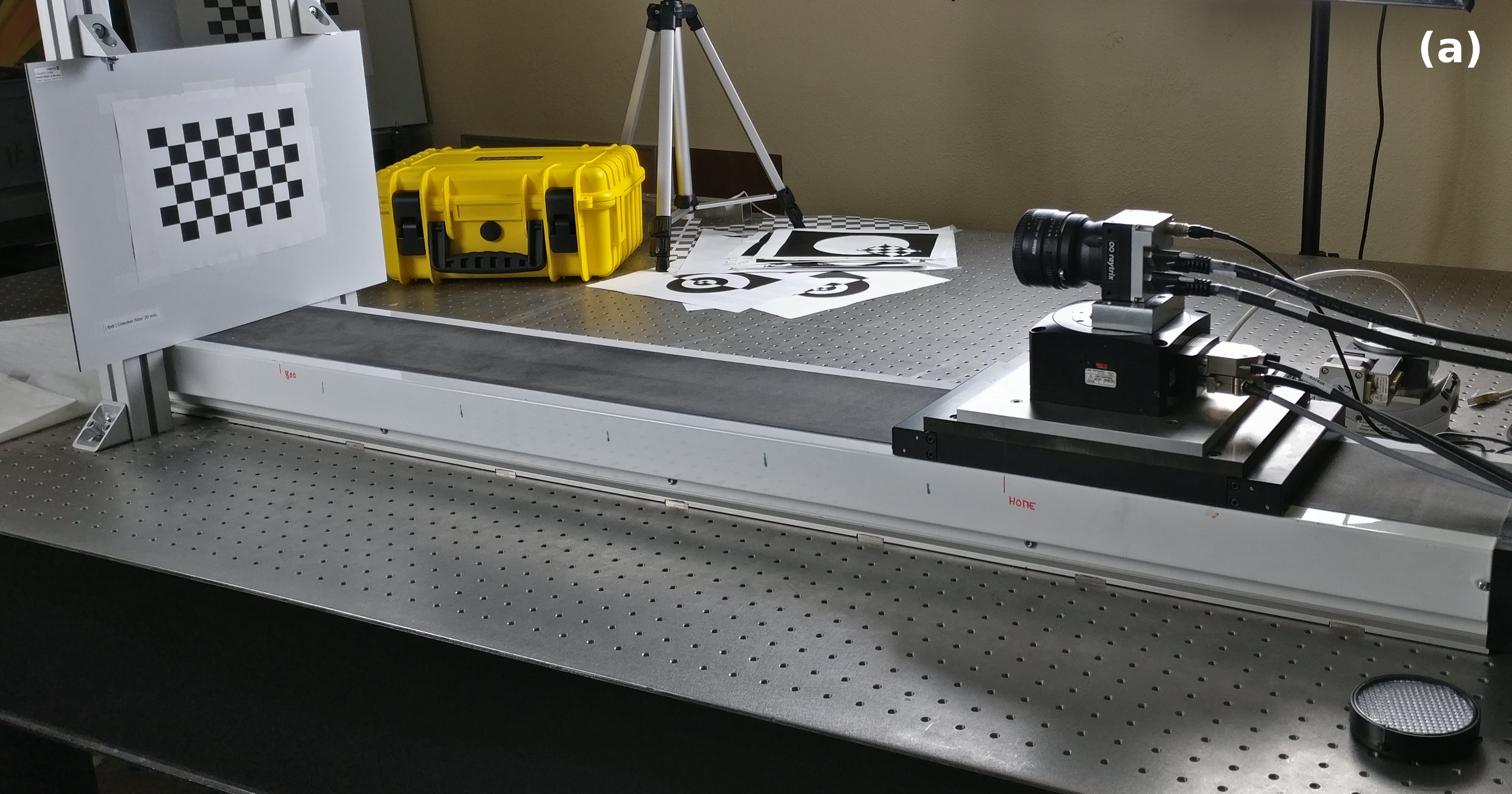}\vspace*{-2.3cm}
		\hspace*{-2cm}\fbox{\includegraphics[width=0.65\linewidth]{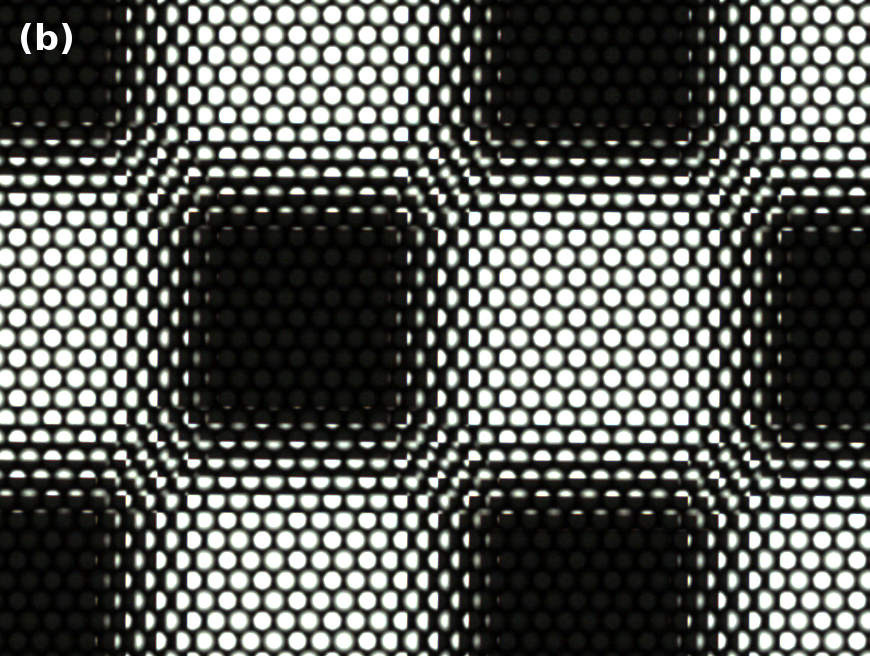}}\hfill\\ %
		\vspace*{-3cm}\fbox{\includegraphics[width=0.35\linewidth]{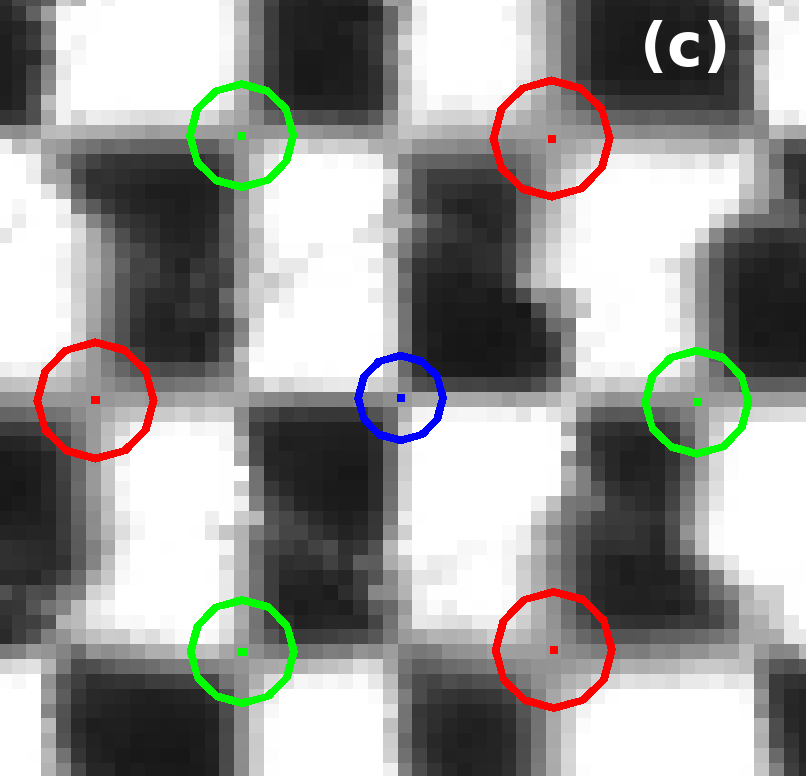}}\hspace*{-4.92cm}\hfill
	\end{center}
	\vspace*{-0.5cm}
	\caption{
		The \texttt{Raytrix R12} multi-focus plenoptic camera used in our experimental setup (a), along with a raw image of a checkerboard calibration target (b). 
		The image is composed of several micro-images with different blurred levels and arranged in an hexagonal grid.
		In each micro-image, our new \acrfull{BAP} feature is illustrated by its center and its blur radius (c).
	}
	\label{fig:mfpc}
	\vspace*{-0.5cm}
\end{figure}

	The purpose of an imaging system is to map incoming light rays from the scene onto pixels of photo-sensitive detector. 
	The \gls{radiance} of a light ray is given by the \textit{plenoptic function} $\lf{\vect{x}, \vect{\theta}, \lambda, \tau}$, introduced by Adelson \etal~\cite{Adelson1991}, where 
	$\vect{x} \in \Reals^3$ is the \textit{spatial} position of observation,
	$\vect{\theta} \in \Reals^2$ is the \textit{angular} direction of observation,
	$\lambda$ is the wavelength of the light 
	and $\tau$ is the time.
	Conventional cameras capture only one point of view. %
	A \textit{plenoptic camera} is a device that allows to retrieve spatial as well as angular information.

	From \textit{Lumigraph} \cite{Lippmann1911b} to commercial \textit{plenoptic cameras} \cite{Ng2005b,Perwass2010c}, several designs have been proposed. %
	This paper focuses on plenoptic cameras based on a \gls{MLA} placed between the main lens and the photo-sensitive sensor (see \autoref{fig:model}).
	The specific design of such a camera allows to multiplex both types of information onto the sensor in the form of a \gls{MIA} (see \autoref{fig:mfpc} (b)), but implies a trade-off between the angular and spatial resolutions~\cite{Georgiev2006,Levin2008b,Georgiev2009e}.
	It is balanced according to the \gls{MLA} position with respect to the main lens focal plane
	(\ie, \textit{focused}~\cite{Perwass2010c,Georgiev2012} and \textit{unfocused}~\cite{Ng2005b} configurations).
	The mapping of incoming light rays from the scene onto pixels can be expressed as a function of the camera model.
	Classical cameras are usually modeled as pinhole or thin lens. 
	Due to the complexity of plenoptic camera's design, the used models are usually high dimensional.
	Specific calibration methods have to be developed to retrieve the intrinsic parameters of these models.

\subsection{Related Work}

\paragraph{Unfocused plenoptic camera calibration.}
In this configuration, light rays are focused by the \gls{MLA} on the sensor plane.
The calibration of unfocused plenoptic camera \cite{Ng2005b} has been widely studied in the literature~\cite{Dansereau2013c,Bok2014,Shi2016,Shi2019,Hahne2018,Zhou2019}.
Most approaches rely on a thin-lens model for the main lens and an array of pinholes for the micro-lenses.
Most of them require reconstructed images to extract features,
and limit their model to the unfocused configuration, \ie, setting the micro-lens \gls{focal} at the distance \gls{MLA}-sensor.
Therefore those models cannot be directly extended to the focused or multi-focus plenoptic camera.

\paragraph{Focused plenoptic camera calibration.}
With the arrival of commercial focused plenoptic cameras~\cite{Lumsdaine2009b,Perwass2010c}, new calibration methods have been proposed.
Based on~\cite{Johannsen2013b}, Heinze \etal \cite{Heinze2016b} have developed a new projection model and a metric calibration procedure which is incorporated in the \texttt{RxLive} software of \texttt{Raytrix GmbH}.
Other calibration methods and models have been proposed, either to overcome the fragility of the initialization~\cite{Strobl2016}, or to model more finely the camera parameters using depth information~\cite{Zeller2014,Zeller2016h}.
O'Brien \etal~\cite{OBrien2018} introduced a new 3D feature called plenoptic disc %
and defined by its center and its radius.
Nevertheless, all previous methods rely on reconstructed images meaning that they introduce error in the reconstruction step as well as in the calibration process.

To overcome this problem, several calibration methods~\cite{Zhang2016a,Zhang2016,Sun2016,Noury2017b,Bok2017,Nousias2017,Wang2018} have been proposed using only raw plenoptic images.
In particular, features extraction in raw micro-images has been studied in~\cite{Bok2017,Nousias2017,Noury2017b} achieving improved performance through automation and accurate identification of feature correspondences. 
However, most of the methods rely on simplified models for optic elements: the \gls{MLA} is modeled as a pinholes array making it impossible to retrieve the \glspl{focal}, or the \gls{MLA} misalignment is not considered.
Some do not consider distortions \cite{Zhang2016a,Nousias2017,Wang2018} or restrict themselves to the focused case~\cite{Zhang2016a,Zhang2016,Noury2017b}.

Finally, few have considered the multi-focus case~\cite{Heinze2016b,Bok2017,Nousias2017,Wang2018} but dealt with it in separate processes, leading to different intrinsic and extrinsic parameters according to the type of micro-lenses. 
\subsection{Contributions}
\begin{figure}[t]
	\begin{center}\vspace*{-2mm}
		\def\tkzscl{0.75}
		\input{figures/tex/notations-all}
	\end{center}\vspace*{-0.9cm}
	\caption{Focused Plenoptic Camera model in Galilean configuration (\ie, the main lens focuses behind the sensor) with the notations used in this paper.}
	\label{fig:notations}\label{fig:model}\vspace*{-0.6cm}
\end{figure}
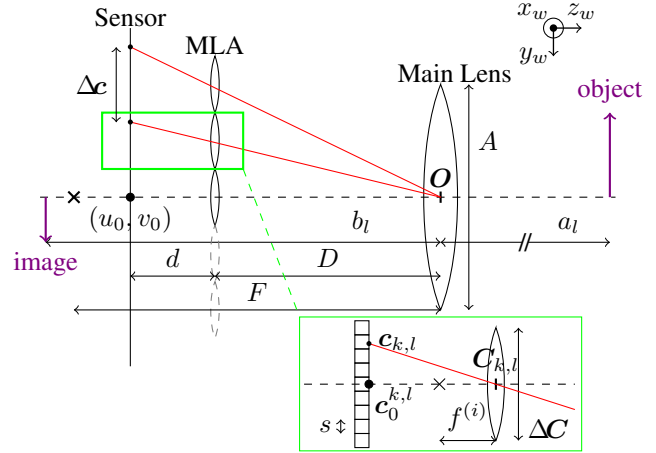

This paper focuses on the calibration of micro-lenses-based \gls{MFPC}. %
To the best of our knowledge, this is the first method proposing a single optimization process that retrieves intrinsic and extrinsic parameters of a \gls{MFPC} directly from raw plenoptic images.
The main contributions are the following:\vspace*{-0.5em}
\begin{itemize}
	\item
	We present 
		a new \gls{BAP} feature defined in raw image space 
		that enables us to handle the multi-focus case.\vspace*{-0.5em}
	\item 
	We introduce
		a new pre-calibration step using \gls{BAP} features from white images to provide a robust initial estimate of internal parameters.\vspace*{-0.5em}
	\item
	We propose
		a new reprojection error function exploiting \gls{BAP} features to refine a more complete model, including in particular the multiple micro-lenses \glspl{focal}.
		Our checkerboard-based calibration is conducted in a single optimization process.
\end{itemize}
A visual overview of our method is given in \autoref{fig:detecabstract}.
The remainder of this paper is organized as follows:
first, the camera model and \gls{BAP} feature are presented in \autoref{sec:cameramodel}.
The proposed pre-calibration step is explained in \autoref{sec:precalib}.
Then, the feature detection is detailed in \autoref{sec:features} and the calibration process in \autoref{sec:calib}.
Finally, our results are presented and discussed in \autoref{sec:results}.
The notations used in this paper are shown in \autoref{fig:notations}.
Pixel counterparts of metric values are denoted in lower-case Greek letters.
\section{Camera model and \gls{BAP} feature}\label{sec:cameramodel}

\subsection{Multi-focus Plenoptic Camera} 
We consider multi-focus plenoptic cameras as described in~\cite{Georgiev2012,Perwass2010c}.
The main lens, modeled as a thin-lens, maps object point to virtual point behind (\resp in front of) the image sensor in Galilean (\resp Keplerian) configuration.
Therefore, the \gls{MLA} consists of $\mltypenb$ different lens types with focal lengths $\focal\mltype{i}, i \in \set{1,\dots, \mltypenb}$
which are focused on $\mltypenb$ different planes %
behind the image sensor. %
This multi-focus setup corresponds to the \texttt{Raytrix} camera system described in~\cite{Perwass2010c} when $\mltypenb=3$.	
The micro-lenses are modeled as thin-lenses allowing to take into account blur in the micro-image.
Our model takes into account the \gls{MLA} misalignment, freeing all six degrees of freedom.

The tilt of the main lens is included in the distortion model and we make the hypothesis that the main lens plane $\plane\flens$ is parallel to the sensor plane $\plane\fsensor$.
Furthermore, we choose the main lens frame as our camera reference frame, with $\lenscenter$ being the origin, the $z$-axis coinciding with the optical axis and pointing outside the camera, and the $y$-axis pointing downwards. %
Only distortions of the main lens are considered.
We use the model of Brown-Conrady~\cite{Brown1966,Conrady1919}
with three coefficients for the radial component
and two for the tangential.
Furthermore, we take into account the deviation of the image center and the optical center for each micro-lens because it tends to cause inaccuracy in decoded light field. 
Therefore, the principal point $\mlpp$ of the micro-lens indexed by $\left(\mlindexes\right)$ is given by
\begin{equation}\label{eq:mlprincipalpoint}
\mlpp = \bbm\mlppx\\\mlppy\ebm = \frac{\dist}{\Dist+\dist}\left(\bbm \principalpointx \\ \principalpointy \ebm - \micenter_{\mlindexes}\right) + \micenter_{\mlindexes}\tcomma
\end{equation}
where $\micenter_{\mlindexes}$ is the center in pixels of the micro-image $\left(\mlindexes\right)$,
$\bbm\principalpointx& \principalpointy\ebm\T$ is the main lens principal point,
$\dist$ is the distance \gls{MLA}-sensor
and
$\Dist$ is the distance main lens-\gls{MLA},
as illustrated in~\autoref{fig:notations}.

Finally, each micro-lens produces a micro-image onto the sensor. 
The set of these micro-images has the same structural organization as the \gls{MLA}, \ie, in our case an hexagonal grid, alternating between each type of micro-lens. 
The data can therefore be interpreted as an array of micro-images, called by analogy the \gls{MIA}.
The \gls{MIA} coordinates are expressed in image space.
Let $\miinterdist$ be the pixel distance between two arbitrary consecutive micro-images centers $\micenter_{\mlindexes}$.
With $\pixelsize$ the metric size of a pixel, let $\miinterdistmm = \pixelsize \cdot \miinterdist$ be its metric value,
and $\mlinterdist$ be the metric distance between the two corresponding micro-lenses centers $\mlcenter_{\mlindexes}$.
From similar triangles, the ratio between them is given by
\begin{equation}\label{eq:mlinterdist}
\frac{\mlinterdist}{\miinterdistmm} = \frac{\distlensmla}{\distmlasensor+\distlensmla} \Longleftrightarrow \mlinterdist = \miinterdistmm \cdot \frac{\distlensmla}{\distmlasensor+\distlensmla}\tdot
\end{equation}
We make the hypothesis that $\mlinterdist$ is equal to the micro-lens diameter.
Since $\dist \ll \Dist$, we can make the following approximation:
\begin{equation}\label{eq:mldiameterapprox2}
\frac{\Dist}{\Dist+\dist} = \lambda \approx 1 \Longrightarrow \mldiameter = \miinterdistmm\cdot\frac{\Dist}{\Dist+\dist} \approx \miinterdistmm\tdot
\end{equation}
This approximation will be validated in the experiments.

\subsection{\gls{BAP} feature and projection model}

Using a camera with a circular aperture, the blurred image of a point on the image detector is circular in shape and is called the \textit{blur circle}. 
From similar triangles and from the thin-lens equation, the signed blur radius of a point in an image can be expressed as 
\begin{equation}\label{eq:blurradius}
\blurradiuspix = \frac{1}{\pixelsize}\cdot\frac{\aperture}{2}\dist\left(\frac{1}{\focal} - \frac{1}{\distobj}- \frac{1}{\dist}\right)\tcomma
\end{equation}
with 
$\pixelsize$ the size of a pixel,
$\dist$ the distance between the considered lens and the sensor,
$\aperture$ the \gls{aperture} of this lens,
$\focal$ its \gls{focal},
and
$\distobj$ the distance of the object from the lens.

This radius appears at different levels in the camera projection: %
in the blur introduced by the thin-lens model of the micro-lenses and during the formation of the micro-image while taking a white image.
To leverage blur information, we introduce a new \acrfull{BAP} feature characterized by its center and its radius, \ie, $\feature = \left(u, v, \blurradiuspix\right)$.
Therefore, our complete plenoptic camera model allows us to link a scene point $\point\fworld$ to our new \gls{BAP} feature $\feature$ through each micro-lens $\left(\mlindexes\right)$
\begin{equation}\label{eq:completemodel}
\bbm u \\ v \\ \blurradiuspix \\ 1\ebm \propto
\mat{\mathcal{P}}\left(i,\mlindexes\right)
\cdot 	\poseMLkl
\cdot 	\distor{
	\Kthinlens\left(\Focal\right)
	\cdot 	\poseLens
	\cdot 	\point\fworld
}\tcomma%
\end{equation}
where
$\mat{\mathcal{P}}\left(i,\mlindexes\right)$ is the blur aware plenoptic projection matrix through the micro-lens $\left(\mlindexes\right)$ of type $i$, and computed as
\begin{align}
&\mat{\mathcal{P}}\left(i,\mlindexes\right) = \mat{P}\left(\mlindexes\right)\cdot \Kthinlens\left(\focal\mltype{i}\right) \\
=& \bbm 
{\dist}/{\pixelsize} & 0 & \principalpointx^{\mlindexes} & 0 \\
0 & {\dist}/{\pixelsize} & \principalpointy^{\mlindexes} & 0 \\
0 & 0 & \pixelsize\frac{\mlinterdist}{2} & -\pixelsize\frac{\mlinterdist}{2}\dist \\
0 & 0 & -1 & 0		
\ebm 
\bbm 
1 & 0 & 0 & 0 \\
0 & 1 & 0 & 0 \\
0 & 0 & 1 & 0\\
0 & 0 & -1/\focal\mltype{i} & 1	
\ebm\tdot\nonumber
\end{align}
$\mat{P}\left(\mlindexes\right)$ is a matrix that projects the 3D point onto the sensor.
$\Kthinlens\left(\focal\right)$ is the thin-lens projection matrix for the given \gls{focal}.
$\poseLens$ is the pose of the main lens with respect to the world frame 
and $\poseMLkl$ is the pose of the micro-lens $\left(\mlindexes\right)$ expressed in the camera frame.
The function $\funarg{\Distor}{\cdot}$ models the lateral distortion.

Finally, the projection model defined in \autoref{eq:completemodel} consists of a set $\Xi$ of $\left(16+\mltypenb\right)$	
intrinsic parameters to be optimized, including
the main lens \gls{focal} $\Focal$ 
and its 5 lateral distortion parameters, %
the sensor translation, encoded in $\left(\principalpointx, \principalpointy\right)$ and $\dist$,
the \gls{MLA} misalignment, \ie, 3 rotations $\left(\theta_x, \theta_y, \theta_z\right)$ and 3 translations $\left(t_x, t_y, \Dist\right)$,
the micro-lens inter-distance $\mlinterdist$,
and the $\mltypenb$ micro-lens \glspl{focal}  $\focal\mltype{i}$.
\section{Pre-calibration using raw white images}\label{sec:precalib}

\begin{figure}[t]
	\centering
	\includegraphics[width=1\linewidth]{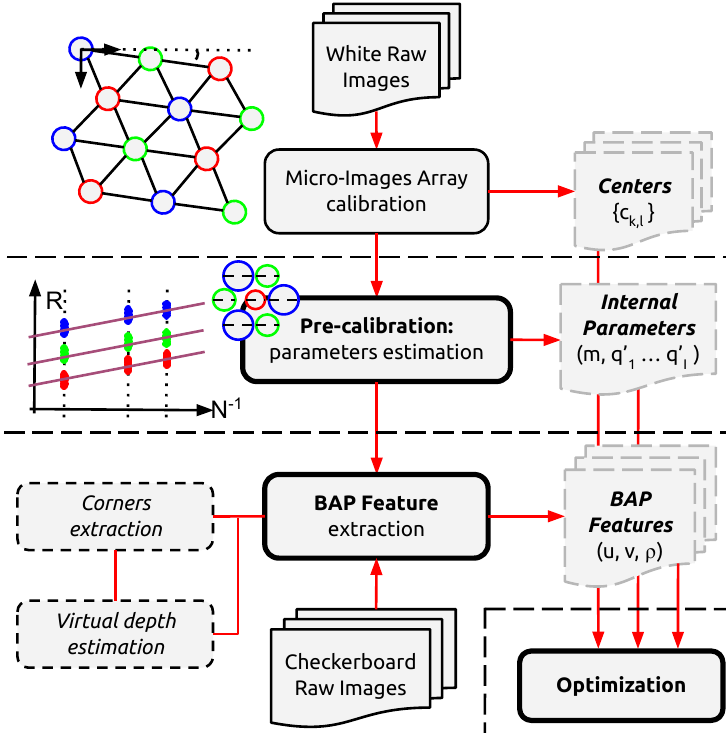}%
	\caption{Overview of our proposed method with the pre-calibration step and the detection of \gls{BAP} features that are used in the non-linear optimization process. %
	}
	\label{fig:detecabstract}
\end{figure}

Drawing inspiration from \textit{depth from defocus} theory~\cite{Subbarao1994}, 
we leverage blur information to estimate parameters (\eg, here our blur radius) by varying some other parameters (\eg, the \gls{focal}, the \gls{aperture}, etc.) in combination with known (\ie, fixed or measured) parameters.
For instance, when taking a white picture with a controlled \gls{aperture}, each type of micro-lens produces a \acrfull{MI} with a specific size and intensity, providing a way to distinguish between them.
In the following all distances are given with reference to the \gls{MLA} plane.
Distances are signed according to the following convention: $\focal$ is positive when the lens is convergent; distances are positive when the point is real, and negative when virtual.

\subsection{Micro-image radius derivation}

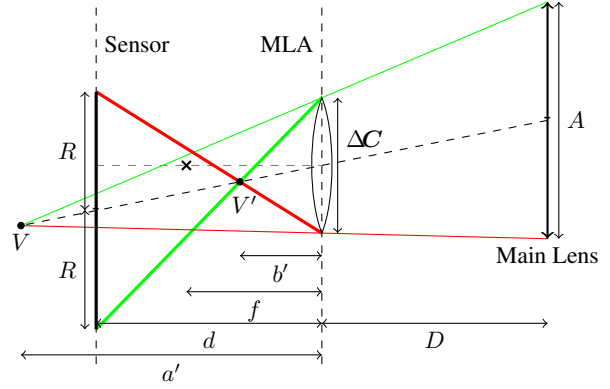
\begin{figure}[t]
	\begin{center}
		\def\tkzscl{0.6}
		\input{figures/tex/miradius}
	\end{center} \vspace*{-0.5cm}
	\caption{%
		Formation of a micro-image with its radius $\miradiusmm$ through a micro-lens while taking a white image at an \gls{aperture} $\aperture$.
		The point $V$ is the vertex of the cone passing by the main lens and the considered micro-lens.  
		$V'$ is the image of $V$ by the micro-lens and $\miradiusmm$ is the radius of its blur circle.	
	}
	\label{fig:blurradius}\vspace*{-0.4cm}
\end{figure}

Taking a white image is equivalent for the micro-lenses to image a white uniform object of diameter $\aperture$ at a distance $\Dist$.
This is illustrated in~\autoref{fig:blurradius}.
We relate the \gls{MI} radius to the plenoptic camera parameters.
From optics geometry, %
the image of this object, \ie the resulting \gls{MI}, is equivalent to the image of an imaginary point constructed as the vertex of the cone passing through the main lens and the considered micro-lens (noted $V$ in~\autoref{fig:blurradius}).
Let $\distobj'$ be the distance of this point from the \gls{MLA} plane, 
given from similar triangles and \autoref{eq:mlinterdist} by
\begin{equation}
\distobj' = -\Dist\frac{\mldiameter}{\aperture-\mldiameter} 
= - {\Dist}\left(\aperture\left(\frac{\dist+\Dist}{\miinterdistmm\Dist}\right) - 1\right)^{-1}\tcomma
\end{equation}
with 
$\aperture$ the main lens \gls{aperture}.
Note the minus sign is due to the fact that the imaginary point is always formed behind the \gls{MLA} plane at a distance $\distobj'$, and thus considered as a virtual object for the micro-lenses.
Conceptually, the \gls{MI} formed can be seen as the \textit{blur circle} of this imaginary point.
Therefore, using  \autoref{eq:blurradius}, the metric \gls{MI} radius $\miradiusmm$ is given by
\begin{align}\label{eq:miradiusaperture}
\miradiusmm &= \frac{\mldiameter}{2}\dist\left(\frac{1}{\focal} - \frac{1}{\distobj'} - \frac{1}{d}\right) \nonumber\\ 
&= \aperture\cdot \frac{\dist}{2\Dist} +  \left(\frac{\miinterdistmm\Dist}{\dist+\Dist}\right) \cdot \frac{\dist}{2} \cdot \left(\frac{1}{\focal} - \frac{1}{\Dist} - \frac{1}{d}\right)
\tdot
\end{align}
From the above equation, we see that the radius depends linearly on the \gls{aperture} of the main lens. %
However, the main lens \gls{aperture} cannot be computed directly whereas we have access to the \tfnumber value.
The \tfnumber of an optical system is the ratio of the system's \gls{focal} $\Focal$ to the diameter of the entrance pupil, $\aperture$, given by
$\fnumber = {\Focal}/{\aperture}$.
\noindent Finally, we can express the \gls{MI} radius for each micro-lens \gls{focal} type $i$ as
\begin{equation}\label{eq:miradiusfnumber}
\funarg{\miradiusmm}{\fnumber^{-1}} = m \cdot \fnumber^{-1} + q_i
\end{equation}
with
\begin{equation}
m = \frac{\dist\Focal}{2\Dist} \text{~~~and~~~}
q_i = \frac{1}{\focal\mltype{i}} \cdot \left(\frac{\miinterdistmm\Dist}{\dist+\Dist}\right) \cdot \frac{\dist}{2}  - \frac{\miinterdistmm}{2} \tdot\label{eq:miradiusfnumbermc}
\end{equation}%
Let $q_i'$ be the value obtained by
$q_i' = q_i + {\miinterdistmm}/{2}$.

\subsection{Internal parameters estimation}\label{subsec:internalparam}
The \textit{internal parameters} $\internalparam = \set{m, q'_1, \dots, q'_\mltypenb}$ are used to compute the radius part of the \gls{BAP} feature and to initialize the parameters of the calibration process.
Given several raw white images taken at different \glspl{aperture}, we estimate the coefficients of \autoref{eq:miradiusfnumber} %
for each type of micro-image.
The standard full-stop \tfnumber conventionally indicated on the lens differs from the real \tfnumber calculated	
with the \gls{aperture} value $\aperturevalue$ as
$\fnumber = \sqrt{2^{\aperturevalue}}\tdot$
From raw white images, we are able to measure each \acrfull{MI} radius $\miradiuspix = \abs{\miradiusmm}/\pixelsize$ in \si{pixels} for each distinct \gls{focal} $\focal\mltype{i}$ at a given \gls{aperture}.
Due to \gls{vignetting} effect, the estimation is only conducted on center micro-images which are less sensitive to this effect.
Our method is based on image moments fitting.
It is robust to noise, works under asymmetrical distribution and is easy to use, but needs a parameter $\alpha$ to convert the standard deviation $\sigma$ into a pixel radius $\miradiuspix = \alpha\cdot\sigma$.
We use the second order central moments of the micro-image to construct a covariance matrix.
Finally, we choose $\sigma$ as the square root of the greater eigenvalue of the covariance matrix.
The parameter $\alpha$ is determined such that at least $98\%$ of the distribution is taken into account.
According to the standard normal distribution $Z$-score table, $\alpha$ is picked up in $\range{2.33, 2.37}$.
In our experiments, we set $\alpha = 2.357$.

\begin{figure}[t]
	\begin{center}
		\includegraphics[width=0.66\linewidth]{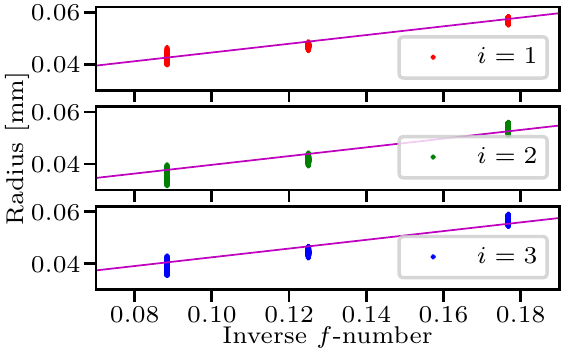}\vspace*{0.5mm}
		\raisebox{0.35\height}{\includegraphics[width=0.32\linewidth]{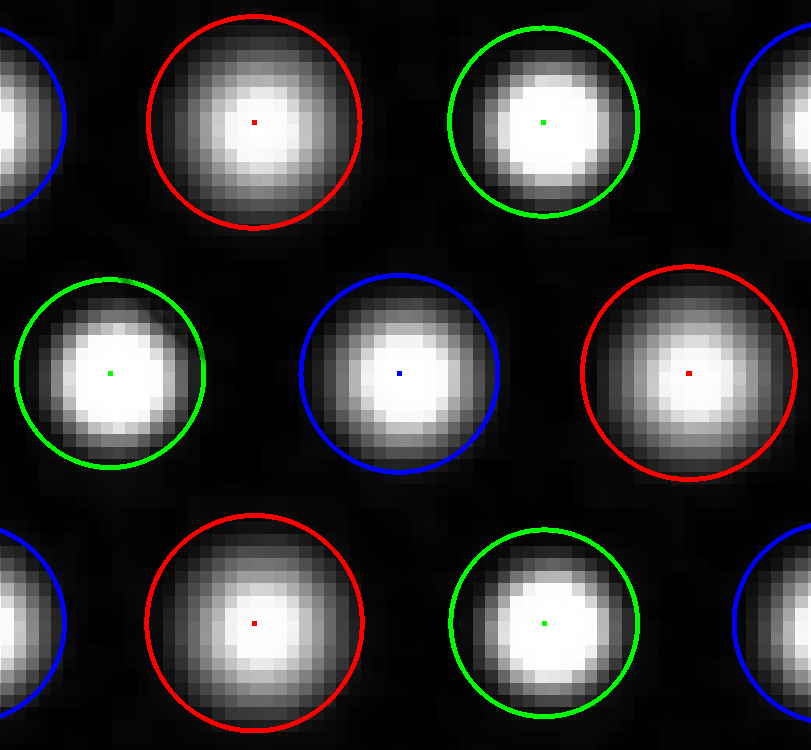}}
		\includegraphics[width=\linewidth]{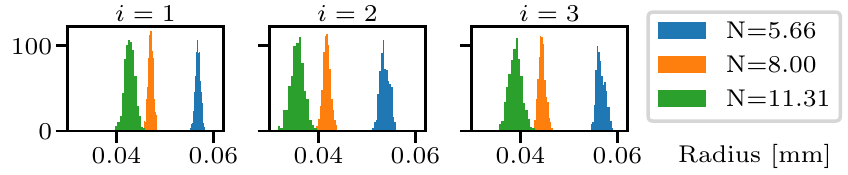}		
	\end{center}
	\vspace*{-0.6cm}
	\caption{
		Micro-image radii as function of the inverse $f$-number with the estimated lines (in \textit{magenta}).
		Each type of micro-lens is identified by its color (type (1) in \textit{red}, type (2) in \textit{green} and type (3) in \textit{blue}) with its computed radius. 	
		For each type an histogram of the radii distribution is given.
	}\label{fig:radii}\vspace*{-2mm}
\end{figure}

Finally, from radii measurements at different $f$-numbers, we estimate the coefficients of \autoref{eq:miradiusfnumber}, $\vect{X} = \set{m, q_1, \dots, q_\mltypenb}$, with a least-square estimation.
\autoref{fig:radii} shows the radii computed in a white image taken with an $f$-number $\fnumber = 8$, the histogram of these radii distributions for each type, and the estimated linear functions.
\section{\gls{BAP} feature detection in raw images}\label{sec:features}

At this point, the internal parameters $\internalparam$, used to express our blur radius $\blurradiuspix$, are available.
The process (illustrated in \autoref{fig:detecabstract}) is divided into three phases:
	1) using a white raw image, the \gls{MIA} is calibrated and micro-image centers are extracted;
	2) checkerboard images are processed to extract corners at position $\left(u,v\right)$;
	and 3) with the internal parameters and the associated virtual depth estimate for each corner, the corresponding \gls{BAP} feature is computed.

\subsection{Blur radius derivation through micro-lens}
To respect the $f$-number matching principle~\cite{Perwass2010c}, we configure the main lens \tfnumber such that the micro-images fully tile the sensor without overlap.
In this configuration the working \tfnumber of the main imaging system and the micro-lens imaging system should match.
The following relation is then verified for at least one type of micro-lens:
\begin{equation}
\frac{\mldiameter}{\distmlasensor} = \frac{\aperture}{\distlensmla} \Longleftrightarrow \frac{1}{\dist}\cdot\left(\frac{\miinterdistmm\Dist}{\dist + \Dist}\right) = \frac{\Focal}{\fnumber\Dist}\tdot
\end{equation}
We consider the general case of measuring an object $\point$ at a distance $\distobj\flens$ from the main lens.
First, $\point$ is projected through the main lens according to the thin lens equation,
${1}/{F} = {1}/{\distobj\flens}+{1}/{\distimg\flens}$, 
resulting in a point $\point'$ at a distance $\distimg\flens$ behind the main lens, 
\ie at a distance 
$\distobj =  \Dist - \distimg\flens$
from the \gls{MLA}. %
The metric radius of the blur circle $\blurradiusmm$ formed on the sensor for a given point $\point'$ at distance $\distobj$ through a micro-lens of type $i$ is expressed as
\begin{align}\label{eq:radiusprefinal}
\blurradiusmm &= \left(\frac{\miinterdistmm\Dist}{\dist+\Dist}\right) \cdot \frac{\dist}{2} \cdot \left(\frac{1}{\focal\mltype{i}} - \frac{1}{\distobj} - \frac{1}{\dist} \right) \nonumber\\
&= 
\underbrace{
	\frac{\miinterdistmm\Dist}{\dist+\Dist} \cdot \frac{\dist}{2} \cdot \frac{1}{\focal\mltype{i}}
}_{= q'_i \text{~\ref{eq:miradiusfnumbermc}}} 
-
\underbrace{
	\frac{\miinterdistmm\Dist}{\dist+\Dist} \cdot \frac{\dist}{2} \cdot \frac{1}{\dist}
}_{= \mlinterdist/2 \text{~\ref{eq:mlinterdist}}}  - 
\underbrace{
	\frac{\miinterdistmm\Dist}{\dist+\Dist}
}_{= \mlinterdist \text{~\ref{eq:mlinterdist}}} \cdot \frac{\dist}{2} \cdot \frac{1}{\distobj}\nonumber\\
&= \left(-\mlinterdist\cdot\frac{\dist}{2}\right) \cdot \frac{1}{\distobj} + \left( q_i'	- \frac{\mlinterdist}{2}\right)\tdot	
\end{align}
In practice, we do not have access to the value of $\mlinterdist$ but we can use the approximation from \autoref{eq:mldiameterapprox2}. %
Moreover, $\distobj$ and $\dist$ cannot be measured in the raw image space, but the virtual depth can.
Virtual depth refers to relative depth value obtained from disparity. 
It is defined as the ratio between the object distance $\distobj$ and the sensor distance $\dist$:
\begin{equation}
\virtdepth = \frac{\distobj}{\dist}\tdot
\end{equation}
The sign convention is reversed for virtual depth computation, \ie distances are negative in front of the \gls{MLA} plane.
If we re-inject the latter in \autoref{eq:radiusprefinal}, taking caution of the sign, we can derive the radius of the \textit{blur circle} of a point $\point'$ at a distance $\distobj$ from the \gls{MLA} by
\begin{equation}\label{eq:radiusfinal}
\begin{aligned}
\blurradiusmm &= \frac{\mlinterdistapprox}{2}\cdot \virtdepth^{-1} + \left( q_i'	- \frac{\mlinterdistapprox}{2}\right)\tdot
\end{aligned}
\end{equation}
This equation allows to express the pixel radius of the blur circle $\blurradiuspix = \blurradiusmm/\pixelsize$ associated to each point having a virtual depth without explicitly evaluating $\aperture, \Dist, \dist, \Focal \tand \focal\mltype{i}$.

\subsection{Features extraction}

First, the \Acrlong{MIA} has to be calibrated. 
We compute the micro-image centers observations $\set{\micenter_{\mlindexes}}$ by intensity centroid method with sub-pixel accuracy~\cite{Thomason2014,Noury2017b,Suliga2018}.
The distance between two micro-image centers $\miinterdist$ is then computed as the optimized edge-length of a fitted 2D grid mesh with a non-linear optimization.
The pixel translation offset in the image coordinates, $(\tau_x,\tau_y)$,
and the rotation around the $\left(-z\right)$-axis, $\vartheta_z$,
are also determined during the optimization process.
From the computed distance and the given size of a pixel, the parameter $\mlinterdistapprox$ is computed.

Secondly, as we based our method on checkerboard calibration pattern, 
we detect corners in raw images using the detector introduced by Noury \etal~\cite{Noury2017b}.
The raw images are devignetted by dividing them by a white raw image taken at the same \gls{aperture}.

With a plenoptic camera, contrarily to a classical camera, a point is projected into more than one observation onto the sensor.
Due to the nature of the spatial distribution of the data, we used the \acrshort{DBSCAN} algorithm~\cite{Ester1996} to identify the clusters.
We then associate each point with its cluster of observations.

Once each cluster is identified, we can compute the virtual depth $\virtdepth$ from disparity.
Given the distance $\Delta\mlcenter_{1-2}$ between the centers of the micro-lenses $\mlcenter_1$ and $\mlcenter_2$, \ie the baseline, and the Euclidean distance $\Delta\imgpt = \abs{\imgpt_1 - \imgpt_2}$ between images of the same point in corresponding micro-images, the virtual depth $\virtdepth$ can be calculated with the intercept theorem:
\begin{equation}\label{eq:virtdepth}
\virtdepth = \frac{\Delta\mlcenter_{1-2}}{\Delta\mlcenter_{1-2} - \Delta\imgpt} 
=  \frac{\eta\mlinterdist}{\eta\mlinterdist - \Delta\imgpt} = \frac{\eta\mlinterdistapprox}{\eta\mlinterdistapprox - \Delta\imgpt}\tcomma
\end{equation}
where $\mlinterdist = \mlinterdistapprox$ is the distance between two consecutive micro-lenses and $\eta \ge 1.0$.
Due to noise in corner detection, we use a median estimator to compute the virtual depth of the cluster taking into account all combinations of point pairs in the disparity estimation.

Finally, from internal parameters $\internalparam$ and with the available virtual depth $\virtdepth$  we can compute the \gls{BAP} features using \autoref{eq:radiusfinal}.
In each frame $n$, for each micro-image $\left(\miindexes\right)$ of type $i$ containing a corner at position $\left(u,v\right)$ in the image, the feature $\feature_{\miindexes}^n$ is given by
\begin{equation}
\feature_{\miindexes}^n = \left(u,v,\rho\right), \twith \rho = \blurradiusmm / \pixelsize\tdot
\end{equation}
In the end, our observations are composed of a set of micro-image centers $\set{\micenter_{\miindexes}}$ and a set of \gls{BAP} features $\set{\feature_{\miindexes}^n}$ allowing us to introduce two reprojection error functions corresponding to each set of features.

\section{Calibration process}\label{sec:calib}

To retrieve the parameters of our camera model, %
we use a calibration process based on non-linear minimization of a reprojection error.
The calibration process is divided into three phases:
	1) the intrinsics are initialized using the internal parameters $\internalparam$;
	2) the initial extrinsics are estimated from the raw checkerboard images;
	and 3) the parameters are refined with a non-linear optimization leveraging our new \gls{BAP} features.
\subsection{Parameters initialization}

Optimization processes are sensitive to initial parameters. %
To avoid falling into local minima during the optimization process, the parameters have to be carefully initialized not too far from the solution.
	First, the camera is initialized in \textit{Keplerian} or \textit{Galilean} configuration. 
	First, given the camera configuration, the internal parameters $\internalparam$, the focus distance $\focusdist$, and from the Eq.~(17) of \cite{Perwass2010c},
	the following parameters are set as
	\begin{equation}
		\distmlasensor \longleftarrow\frac{2m\Focusdist}{F + 4m} \text{~~~~and~~~~} \distlensmla \longleftarrow \Focusdist- 2\distmlasensor\tcomma
	\end{equation}
	where $\Focusdist$ is given by
	\begin{equation}
		\Focusdist = \abs{\frac{\focusdist}{2}\left(1-\sqrt{1\pm 4\frac{\Focal}{\focusdist}}\right)}\tcomma
	\end{equation}
	with positive (\resp negative) sign in \textit{Galilean} (\resp \textit{Keplerian}) configuration.

	The \gls{focal}, and the pixel size $\pixelsize$ are set according to the manufacturer value.
	All distortion coefficients are set to zero.
	The principal point is set as the center of the image.
	The sensor plane is thus set parallel to the main lens plane, with no rotation, at a distance $- \left(\distlensmla+\distmlasensor\right)$.
	Seemingly, the \gls{MLA} plane is set parallel to the main lens plane at a distance $-\distlensmla$. 
	From the pre-computed \gls{MIA} parameters the translation takes into account the offsets $\left(-\pixelsize\tau_x, -\pixelsize\tau_y\right)$ and the rotation around the $z$-axis is initialized with $-\vartheta_z$.
	The micro-lenses inter-distance $\mlinterdist$ is set according to \autoref{eq:mlinterdist}.
	Finally, from internal parameters $\internalparam$, the \glspl{focal} are computed as follows:
	\begin{equation}
	\focal\mltype{i} 	\longleftarrow \frac{\distmlasensor}{2 \cdot q_i'}\cdot\mlinterdist\tdot
	\end{equation}
\subsection{Initial poses estimation}

The camera poses $\set{\Transform\fcam^n}$, \ie, the extrinsic parameters are initialized using the same method as in~\cite{Noury2017b}.
For each cluster of observation, the barycenter is computed. 
Those barycenters can been seen as the projections of the checkerboard corners through the main lens using a standard pinhole model.
For each frame, the pose is then estimated using the Perspective-n-Point (PnP) algorithm~\cite{Kneip2011b}.

\subsection{Non-linear optimization}

The proposed model allows us to optimize all the parameters into one single optimization process.	
We propose a new cost function $\Theta$ taking into account the blur information using our new \gls{BAP} feature. 
The cost is composed of two main terms both expressing errors in the image space: 
1) the blur aware plenoptic reprojection error, where, for each frame $n$, each checkerboard corner $\checkerboardcorner$ is reprojected into the image space through each micro-lens $\left(\mlindexes\right)$ of type $i$ according to the projection model of \autoref{eq:completemodel} and compared to its observations $\observation$; 
and 2) the micro-lens center reprojection error, where, the main lens center $\lenscenter$ is reprojected according to a pinhole model in the image space through each micro-lens $\left(\mlindexes\right)$ and compared to the detected micro-image center $\micenter_{\mlindexes}$.

Let $\mathcal{S} = \set{\Xi, \set{\Transform\fcam^n}}$ be the set of intrinsic $\Xi$ and extrinsic $\set{\Transform\fcam^n}$ parameters to be optimized. 
The cost function $\Theta(\mathcal{S})$ is expressed as 
\begin{equation}
\sum \norm{
	\observation - {\Proj_{\mlindexes}\left(\checkerboardcorner\right)}}^2 
+\sum \norm{\micenter_{\mlindexes} - \Proj_{\mlindexes}\left(\lenscenter\right)}^2\tdot
\end{equation}
The optimization is conducted using the Levenberg-Marquardt algorithm.

\begin{table*}[t]
	\begin{center}\footnotesize
		\begin{tabu}{rl XXX XXX XXX }
			\toprule
			&  &\multicolumn{3}{c}{R12-A ($\focusdist=450 \si{\mm}$)} & \multicolumn{3}{c}{R12-B ($\focusdist=1000 \si{\mm}$)} & \multicolumn{3}{c}{R12-C ($\focusdist=\infty$)}\\
			\midrule
			& Unit & Initial & Ours & \cite{Noury2017b} &  Initial &  Ours  & \cite{Noury2017b} &  Initial &  Ours  & \cite{Noury2017b}\\
			\midrule
			\midrule
			$\Focal$& [\si{\mm}]&
			\num{50} &\nump{3}{49.719920861383585}  &\nump{3}{54.887986627318568}&
			\num{50} &\nump{3}{50.046961426391881} &\nump{3}{51.262198508168005}&
			\num{50} &\nump{3}{50.010806410541306} &\nump{3}{53.321663056716744}\\
			\midrule
			$\Dist$& [\si{\mm}] &
			\nump{2}{56.6576} &\nump{3}{56.696345853466113} &\nump{3}{62.424955873589937}&
			\nump{2}{52.1127} &\nump{3}{52.124786681447532} &\nump{3}{53.296254614647317}&
			\nump{2}{49.3841} &\nump{3}{49.384262495325721} &\nump{3}{52.37898088844485} \\
			$\mlinterdist$ &[\si{\micro\m}] &
			\nump{2}{127.505} &\nump{2}{127.4531794822808} &\nump{2}{127.38116992337872}& 
			\nump{2}{127.47} &\nump{2}{127.44568112225776} &\nump{2}{127.40347541080915}& 
			\nump{2}{127.537} &\nump{2}{127.49890099513844} &\nump{2}{127.41756137206919}\\
			$\focal\mltype{1}$ &[\si{\micro\m}]&
			\nump{2}{578.154} &\nump{2}{577.96617832613306} &-& 
			\nump{2}{581.102} &\nump{2}{580.48321597799135} &-& 
			\nump{2}{554.348} &\nump{2}{556.0911821861525} &-\\
			$\focal\mltype{2}$ &[\si{\micro\m}]&
			\nump{2}{504.456} &\nump{2}{505.20895565925383} &-& 
			\nump{2}{503.958} &\nump{2}{504.32539574443513} &-& 
			\nump{2}{475.984} &\nump{2}{479.02576418863141} &-\\
			$\focal\mltype{3}$ &[\si{\micro\m}]&
			\nump{2}{551.667} &\nump{2}{551.79022424696988} &-& 
			\nump{2}{546.393} &\nump{2}{546.37020429192307} &-& 
			\nump{2}{518.976} &\nump{2}{521.32733597299463} &-\\
			\midrule
			$\principalpointx$ &[\si{pix}]&
			$2039$ & $2042.55$ &$2289.83$ &
			$2039$ & $1790.94$ &$1759.29$ & 
			$2039$ & $1661.95$  &$1487.2$\\			
			$\principalpointy$ &[\si{pix}]&
			$1533$ & $1556.29$ &$1528.24$ &
			$1533$ & $1900.19$ &$1934.87$ & 
			$1533$ & $1726.91$  &$1913.81$\\
			$\distmlasensor$& [\si{\micro\m}]&
			\nump{2}{318.632} &\nump{2}{325.242171}  &\nump{2}{402.323545}& 
			\nump{2}{336.842} &\nump{2}{336.261113}  &\nump{2}{363.173728}&
			\nump{2}{307.929} &\nump{2}{312.617105} &\nump{2}{367.402085} \\
			\bottomrule
		\end{tabu}
	\end{center}\vspace*{-4mm}
	\caption{
		Initial intrinsic parameters for each dataset along with the optimized parameters obtained by our method and with the method of \cite{Noury2017b}.
		Some parameters are omitted for compactness.
	}\label{tab:intrinsics}\label{tab:initials}\vspace*{-4mm}
\end{table*}

\section{Experiments and Results}\label{sec:results}

We evaluate our calibration model quantitatively in a controlled environment and qualitatively when ground truth is not available.
\subsection{Experimental setup}

For all experiments we used a \texttt{Raytrix R12} color 3D-light-field-camera, with a \gls{MLA} of F/2.4 \gls{aperture}.
The mounted lens is a \texttt{Nikon AF Nikkor F/1.8D} with a \num{50}\si{\milli\meter} \gls{focal}.
The \gls{MLA} organization is hexagonal, and composed of $176\times152$ (width $\times$ height) micro-lenses with $\mltypenb=3$ different types.
The sensor is a \texttt{Basler beA4000-62KC} with a pixel size of $\pixelsize= 0.0055 \si{\milli\meter}$.
The raw image resolution is $4080\times3068$.

\vspace*{-2mm}
\paragraph{Datasets.}
We calibrate our camera for three different focus distance configurations $\focusdist$%
and build three corresponding datasets: 
	R12-A for $\focusdist = 450$ \si{\mm},
	R12-B for $\focusdist = 1000$ \si{\mm}, 
	and R12-C for $\focusdist = \infty$.
Each dataset is composed of:\vspace*{-0.5em}
\begin{itemize}
	\item white raw plenoptic images acquired at different \glspl{aperture} ($\fnumber \in \set{4, 5.66, 8, 11.31, 16}$) with augmented gain to ease circle detection in the pre-calibration step,\vspace*{-0.5em}
	\item free-hand calibration targets acquired at various poses (in distance and orientation), separated into two subsets, one for the calibration process and the other for the qualitative evaluation,\vspace*{-0.5em}
	\item a white raw plenoptic image acquired in the same luminosity condition and with the same \gls{aperture} as in the calibration targets acquisition,\vspace*{-0.5em}
	\item and calibration targets acquired with a controlled translation motion for quantitative evaluation, along with the depth maps computed by the \texttt{Raytrix} software (\texttt{RxLive}  v4.0.50.2).
\end{itemize}
We use a $9\times5$ of $10\si{\milli\meter}$ side checkerboard for R12-A, 
	a $8\times5$ of $20\si{\milli\meter}$ for R12-B, %
	and a $7\times5$ of $30\si{\milli\meter}$ for R12-C. %
Datasets and our source code are publicly available\footnote{
	\url{https://github.com/comsee-research}
}.

\paragraph{Free-hand camera calibration.}
The white raw plenoptic image is used for devignetting other raw images and for computing micro-images centers. 
From the set of calibration targets images, \gls{BAP} features are extracted, 
and camera intrinsic and extrinsic parameters are then computed using our non-linear optimization process. %

\vspace*{-3.5mm}
\paragraph{Controlled environment evaluation.}
In our experimental setup (see \autoref{fig:mfpc}), the camera is mounted on a linear motion table with micro-metric precision.
We acquired several images with known relative motion between each frame.
Therefore, we are able to quantitatively evaluate the estimated displacement from the extrinsic parameters with respect to the ground truth.
The extrinsics are computed with the intrinsics estimated from the free-hand calibration.
We compared our  computed relative depth to those obtained by the \texttt{RxLive} software.

\vspace*{-3.5mm}
\paragraph{Qualitative evaluation.}
When no ground truth is available, %
we evaluate qualitatively our parameters on the evaluation subset by estimating the reprojection error using the previously computed intrinsics.
We use the \gls{RMSE} as our metric to evaluate the reprojection, individually on the corner reprojection and the blur radius reprojection.

\vspace*{-3.5mm}
\paragraph{Comparison.}
Since our model is close to~\cite{Noury2017b}, we compare our intrinsics with the ones obtained under their pinhole assumption using only corner reprojection error and with the same initial parameters.
We also provide the calibration parameters obtained from the \texttt{RxLive} software.

\subsection{Results}

\paragraph{Internal parameters.}

Note that the pixel \gls{MI} radius is given by $\miradiuspix = \abs{\miradiusmm}/\pixelsize$, 
and $\miradiusmm$ is either positive if formed after the rays inversion (as in \autoref{fig:blurradius}), or negative if before.
With our camera $\focal\mltype{i}>\distmlasensor$~\cite{Heinze2014}, so $\miradiusmm < 0$ implying that $m$ and $c_i$ are also negative.
In practice, it means we use the value $-\miradiuspix$ in the estimation process.

In our experiments, we set $\lambda = 0.9931$. 
We verified that the error introduced by this approximation is less than the metric size of a pixel (\ie, less than \SI{1}{\percent} of its optimized value).
\autoref{fig:radii} shows the estimated lines (see \autoref{eq:miradiusfnumber}), with a cropped white image taken at $\fnumber = 8$, where each type of micro-lens is identified with its radius.
An histogram of the radii distribution is also given for each type.
From the radii measurements, the computed internal parameters are estimated and given for each dataset in \autoref{tab:internals}. 

\begin{table}[h]%
	\begin{center}\footnotesize
		\begin{tabu}{c X X X}
			\toprule
			 & R12-A & R12-B  & R12-C\\
			\midrule
			\midrule
			$\miinterdistmm$ &\nump{3}{128.22163189288707} &\nump{3}{128.29345695445207} &\nump{3}{128.3330652796125}\\
			$m$&\nump{3}{-140.59554040431976} &\nump{3}{-159.56199169158936} &\nump{3}{-155.97544610500336} \\
			$q'_1$&\nump{3}{35.135191355} &\nump{3}{036.489007285581368} &\nump{3}{035.442640815322377}\\
			$q'_2$&\nump{3}{40.268205527646997} &\nump{3}{042.07459207509591} &\nump{3}{041.277697120254017}\\
			$q'_3$&\nump{3}{36.822146226151445} &\nump{3}{038.806929475413102} &\nump{3}{037.858320210521199}\\
			\midrule
			$m^*$&\nump{3}{-142.610734} &\nump{3}{-161.428449} &\nump{3}{-158.291657} \\
			$\epsilon_m$&1.41\% &1.16\% &1.52\% \\
			\bottomrule
		\end{tabu}
	\end{center}\vspace*{-4mm}
	\caption{
		Internal parameters (in \si{\micro\meter}) computed during the pre-calibration step for each dataset.
		The expected value for $m$ is given by $m^*$ and the relative error $\epsilon_m = (m^*-m)/m^*$ is computed.
	}\label{tab:internals}\vspace*{-2.2mm}
\end{table}
As expected, the internal parameters $\miinterdistmm$ and $m$ are different for each dataset, as $\Dist$ and $\miinterdist$ vary with the focus distance $\focusdist$, whereas
the $q'_i$ values are close for each dataset.
Using intrinsics, we compare the expected coefficient $m^*$ (from the closed form in \autoref{eq:miradiusfnumbermc} with the optimized parameters) with its calculated $m$ value.
The mean error over all datasets is $\bar{\epsilon}_{m} = 1.36\%$ which is smaller than a pixel.

\vspace*{-3mm}
\paragraph{Free-hand camera calibration.}
The initial parameters for each dataset are given in \autoref{tab:initials} along with the optimized parameters obtained from our calibration and from the method in~\cite{Noury2017b}.
Some parameters are omitted for compactness (\ie, distortions coefficient and \gls{MLA} rotations which are negligible).
The main lens \glspl{focal} obtained with the proprietary \texttt{RxLive} software are: 
$\Focal_{\focusdist=450} = 47.709$ \si{\mm}, $\Focal_{\focusdist=1000} = 50.8942$ \si{\mm}, and $\Focal_{\focusdist=\infty} = 51.5635$ \si{\mm}.

With our method and \cite{Noury2017b}, the optimized parameters are close to their initial value, showing that our method provides a good initialization for our optimization process.
The $\Focal$, $\distmlasensor$ and $\mlinterdist$ are consistent across the datasets with our method.
In contrast, the $\Focal$ and $\distmlasensor$ obtained with \cite{Noury2017b} show a larger discrepancy.
This is also the case for the \glspl{focal} obtained by \texttt{RxLive}.

\vspace*{-3mm}
\paragraph{Poses evaluation.}
\autoref{tab:translationerror} presents the relative translations and their errors with respect to the ground truth for the controlled environment experiment.
Even if our relative errors are similar with \cite{Noury2017b}, we are able to retrieve more parameters.
With our method, absolute errors are of the order of the \si{\mm} (%
	R12-A: $0.37\pm0.15$ \si{\mm}, 
	R12-B: $1.66\pm0.58$ \si{\mm} and
	R12-C: $1.38\pm0.85$ \si{\mm}%
) showing that the retrieved scale is coherent.
Averaging over all the datasets, our method presents the smallest relative error, with low discrepancy between datasets, outperforming the estimations of the \texttt{RxLive} software.

\begin{table}[h]
	\begin{center}\footnotesize
		\begin{tabu}{c XX XX XX X}
			\toprule
			&\multicolumn{2}{c}{R12-A}& \multicolumn{2}{c}{R12-B} & \multicolumn{2}{c}{R12-C} & All\\
			\midrule
			Error [\%]& $\bar{\epsilon}_z$ & $\sigma_z$ & $\bar{\epsilon}_z$ & $\sigma_z$& $\bar{\epsilon}_z$ & $\sigma_z$ & $\bar{\epsilon}_z$\\
			\midrule
			\midrule 
			Ours &
			\nump{2}{3.731}  & \nump{2}{1.47808} & %
			\nump{2}{3.31682} &\nump{2}{1.16987} & %
			\nump{2}{2.95057} &\nump{2}{1.35466} & 
			\textbf{3.33} \\
			\cite{Noury2017b} &
			\nump{2}{6.834}  & \nump{2}{1.17128} & %
			\nump{2}{1.155} & \nump{2}{1.05567} & %
			\nump{2}{2.69739} &\nump{2}{0.86113} &
			\nump{2}{3,563333333}\\
			\texttt{RxLive} &
			\nump{2}{4.63289}  & \nump{2}{2.50638} & %
			\nump{2}{4.25715}  & \nump{2}{5.79136}& %
			\nump{2}{11.5188}  & \nump{2}{3.22358} &
			\nump{2}{6,803333333} \\
			\bottomrule
		\end{tabu}
	\end{center}\vspace*{-4mm}
	\caption{
		Relative translation error along the $z$-axis with respect to the ground truth displacement.
		For each dataset, the mean error $\bar{\epsilon}_z$ and its standard deviation $\sigma_z$ are given.
		Results are given for our method and compared with \cite{Noury2017b} and the proprietary software \texttt{RxLive}.
	}\label{tab:translationerror}\vspace*{-6mm}
\end{table}

\vspace*{-2mm}
\paragraph{Reprojection error evaluation.}

For each evaluation dataset, the total squared pixel error is reported with its computed \gls{RMSE} in \autoref{tab:rmse}.
The error is less than \num{1}\si{pix} per feature for each dataset demonstrating that the computed intrinsics are valid and can be generalized to images different from the calibration set.

\begin{table}[h]
	\begin{center}\footnotesize
		\begin{tabu}{c XX XX XX}
			\toprule
				&\multicolumn{2}{c}{R12-A \footnotesize(\#11424)}& \multicolumn{2}{c}{R12-B \footnotesize(\#3200)} & \multicolumn{2}{c}{R12-C \footnotesize(\#9568)} \\
			\midrule
			 	& Total & RMSE & Total & RMSE & Total & RMSE \\
			\midrule
			\midrule 
			$\bar{\epsilon}_{all}$ &
				\nump{2}{8972.91}  & \nump{3}{0.886253} & %
				\nump{2}{1444.98} &\nump{3}{0.671978} &%
				\nump{2}{5065.33} &\nump{3}{0.727601} \\
			$\bar{\epsilon}_{u,v}$ &
				\nump{2}{8908.65}  & \nump{3}{0.883074} &%
				\nump{2}{1345.2} &\nump{3}{0.648362} &%
				\nump{2}{5046.68} &\nump{3}{0.72626} \\
			$\bar{\epsilon}_\rho$ &
				\nump{3}{64.2567}  & \nump{3}{0.0749981} &%
				\nump{3}{99.78} &\nump{3}{0.176582} & %
				\nump{3}{18.6589} &\nump{3}{0.0441604} \\
			\bottomrule
		\end{tabu}
	\end{center}\vspace*{-4mm}
	\caption{
		Reprojection error for each evaluation dataset with their number of observations.
		For each component of the feature, the total squared pixel error is reported with its computed \gls{RMSE}. 
	}\label{tab:rmse}\vspace*{-6mm}
\end{table}

\section{Conclusion}
To calibrate the \acrlong{MFPC}, state-of-the-art methods rely on simplifying hypotheses, on reconstructed data or require separate calibration processes to take into account the multi-focal aspect.
This paper introduces a new pre-calibration step which allows us to compute our new \gls{BAP} feature directly in the raw image space.
We then derive a new projection model and a new reprojection error using this feature.
We propose a single calibration process based on non-linear optimization that enables us to retrieve camera parameters, in particular the micro-lenses \glspl{focal}.
Our calibration method is validated by qualitative experiments and quantitative evaluations.
In the future, we plan to exploit this new feature to improve metric depth estimation.\\
\vspace*{-7mm}\paragraph{Acknowledgments.}
This work has been supported by the AURA Region and the European Union (FEDER) through the MMII project of CPER 2015-2020 MMaSyF challenge.
We thank Charles-Antoine Noury and Adrien Coly for their insightful discussions and their help during the acquisitions.

{\small
	\bibliographystyle{ieee_fullname}
	\bibliography{./misc/bib/references}
}

\end{document}

%% file: misc/preamble/preamble.tex
\usepackage[l2tabu,orthodox]{nag}

\DeclareFontFamily{OT1}{pzc}{}
\DeclareFontShape{OT1}{pzc}{m}{it}{<-> s * [1.10] pzcmi7t}{}
\DeclareMathAlphabet{\mathpzc}{OT1}{pzc}{m}{it}

\usepackage[
pagebackref=true,colorlinks,
breaklinks=true,bookmarks=false
]{hyperref}

\addto\extrasenglish{%

}

\usepackage[acronym,toc,nowarn,
	nogroupskip=false,
	style=index
]{glossaries-extra}

\makeatletter
\newcommand*{\glsplainhyperlink}[2]{%
	\colorlet{currenttext}{.}%
	\colorlet{currentlink}{\@linkcolor}%
	\hypersetup{linkcolor=currenttext}%
	\hyperlink{#1}{#2}%
	\hypersetup{linkcolor=currentlink}%
}%
\let\@glslink\glsplainhyperlink%
\makeatother

\usepackage{siunitx}
\usepackage[all]{nowidow}

\usepackage{lipsum}
\usepackage{listings}
\definecolor{mygreen}{rgb}{0,0.6,0}
\definecolor{mygray}{rgb}{0.5,0.5,0.5}
\definecolor{mymauve}{rgb}{0.58,0,0.82}

\usepackage{multicol}

\usepackage{calc}  
\usepackage{enumitem}

\usepackage{lscape} %

\usepackage{epigraph}

\usepackage{epstopdf}

\usepackage{import}

\usepackage{subcaption}
\usepackage{wrapfig}

\graphicspath{{./figures/}}

\usepackage{booktabs}

\usepackage{tabu}

\usepackage{longtable}
\usepackage{rotating,tabularx}

\usepackage{amssymb,amsfonts,amsmath,amscd}

\makeatletter
\let\oldtheequation\theequation
\renewcommand\tagform@[1]{\maketag@@@{\ignorespaces#1\unskip\@@italiccorr}}
\renewcommand\theequation{(\oldtheequation)}
\makeatother

\usepackage{bm} 

\usepackage{csvsimple}

%% file: misc/notations/maths.tex
\newcommand{\tcomma}{\text{, }}
\newcommand{\tdot}{\text{.}}
\newcommand{\tand}{\text{ and }}

\newcommand{\twith}{\text{ with }}

\newcommand{\Field}[1]{\mathbb{#1}}

	\newcommand{\Reals}{\Field{R}}

\newcommand{\vect}[1]{\bm{#1}}
\newcommand{\mat}[1]{\bm{#1}}
\newcommand{\bbm}{\begin{bmatrix}}
\newcommand{\ebm}{\end{bmatrix}}

\newcommand{\T}{^\top}%

\newcommand{\fun}[1]{\mathrm{#1}}
\newcommand{\funarg}[2]{\fun{#1}\!\left(#2\right)}

\newcommand{\Proj}{\fun{\Pi}}

\newcommand{\set}[1]{\left\lbrace#1\right\rbrace}
\newcommand{\range}[1]{\left[#1\right]}
\newcommand{\norm}[1]{\left\Vert#1\right\Vert}
\newcommand{\abs}[1]{\left\vert#1\right\vert}

\newcommand{\nump}[2]{\num[round-mode=places,round-precision=#1]{#2}}

%% file: misc/notations/notations.tex
\newcommand{\focal}{f}
\newcommand{\Focal}{F}

\newcommand{\aperture}{A}%
\newcommand{\aperturevalue}{\mathrm{A\!V}}

\newcommand{\focusdist}{h}
\newcommand{\Focusdist}{H}
\newcommand{\virtdepth}{\nu}

\newcommand{\fnumber}{N}

\newcommand{\internalparam}{\Omega}

\newcommand{\miradiusmm}{R}
\newcommand{\miradiuspix}{\varrho}
\newcommand{\blurradiusmm}{r}
\newcommand{\blurradiuspix}{\rho}

\newcommand{\distmlasensor}{\dist}
\newcommand{\distlensmla}{\Dist}

\newcommand{\distimg}{b}
\newcommand{\distobj}{a}

\newcommand{\lenscenter}{\vect{O}}

\newcommand{\mlinterdist}{\Delta\!{\mlcenter}}%
\newcommand{\mldiameter}{\mlinterdist}%
\newcommand{\mlcenter}{\vect{C}}
\newcommand{\mlindexes}{k,l}

\newcommand{\miinterdist}{\delta\!\micenter}
\newcommand{\miinterdistmm}{\Delta\!\micenter}
\newcommand{\micenter}{\vect{c}}
\newcommand{\miindexes}{\mlindexes}%

\newcommand{\mlinterdistapprox}{\lambda\miinterdistmm}

\newcommand{\principalpointx}{u_0}
\newcommand{\principalpointy}{v_0}
\newcommand{\Principalpoint}{\left(\principalpointx, \principalpointy\right)}

\newcommand{\mlpp}{\micenter{}^{\mlindexes}_0}
\newcommand{\mlppx}{\principalpointx^{\mlindexes}}
\newcommand{\mlppy}{\principalpointy^{\mlindexes}}

\newcommand{\pixelsize}{s}%

\newcommand{\Distor}{\fun{\varphi}}
\newcommand{\distor}[1]{\funarg{\varphi}{#1}}

\newcommand{\imgpt}{\vect{i}}

\newcommand{\Kthinlens}{\mat{K}\!}

\newcommand{\Transform}{\mat{T}}

\newcommand{\poseMLkl}{\Transform\fmla\!\left(\mlindexes\right)}
\newcommand{\poseLens}{\Transform\fcam}

\newcommand{\fworld}{_w}
\newcommand{\fcam}{_c}
\newcommand{\fmla}{_\mu}
\newcommand{\fsensor}{_s}
\newcommand{\flens}{_l}

\newcommand{\mltype}[1]{{}^{\left(#1\right)}}
\newcommand{\mltypenb}{I}

\newcommand{\plane}{\Pi}

\newcommand{\Lf}{\mathcal{L}}
\newcommand{\lf}[1]{\Lf\left(#1\right)}

\newcommand{\point}{\vect{p}}

\newcommand{\dist}{d}
\newcommand{\Dist}{D}

\newcommand*{\siecle}[1]{%
	\ifnum#1=1%
	\bsc{\romannumeral #1}\textsuperscript{er}~siècle%
	\else%
	\bsc{\romannumeral #1}\textsuperscript{e}~siècle%
	\fi
}
\newcommand{\resp}{\textit{resp.,~}}

\newcommand{\tfnumber}{$f$-number }

\newcommand{\feature}{\vect{p}}
\newcommand{\observation}{\feature{}^{n}_{\miindexes}}
\newcommand{\checkerboardcorner}{\point{}^{n}\fworld}

%% file: misc/notations/glossary.tex
\addto{\captionsfrench}{
   
}

\makenoidxglossaries

\DeclareDocumentCommand{\newdualentry}{ O{} O{} m m m m } {
  \newglossaryentry{gls-#3}{name={#5},text={#5},
    description={#6},#1
  }
  \newacronym[see=#2]{#3}{#4}{#5\glsadd{gls-#3}}
}

\newdualentry{SLAM} %
	{SLAM}            %
	{\iflanguage{french}{localisation et cartographie simultanées}{Simultaneous Localization and Mapping}}  %
	{\iflanguage{french}
		{(en anglais, \textit{Simultaneous Localization And Mapping} (SLAM)), est le problème de construction ou de mise à jour de carte d'un environnement inconnu tout en localisant simultanément un agent à l'intérieur de celle-ci.} 
		{is the computational problem of constructing or updating a map of an unknown environment while simultaneously keeping track of an agent's location within it.}
	}%
  
\newdualentry{FOV} %
	{FoV}            %
	{\iflanguage{french}{angle de champ}{field of view}}  %
	{\iflanguage{french}
		{(en anglais, \textit{Field of View}), est l'angle que va pouvoir capter un dispositif optique. Il s'agit de la partie de l'espace qu'est capable d'être restituée sur le détecteur.} 
		{is the extent of the observable world that is seen at any given moment. In the case of optical instruments or sensors it is a solid angle through which a detector is sensitive to electromagnetic radiation.}
	}%
  
\newdualentry{LF} %
	{LF}            %
	{\iflanguage{french}{champ de lumière}{light field}} %
	{\iflanguage{french}
		{(en anglais, \textit{light field}), est une fonction qui décrit la quantité de lumière circulant dans toutes les directions à partir de tout les points de l'espace. Cet espace de tout les rayons de lumière possible est donnée par une fonction plénoptique et la magnitude de chaque rayon est donné par la \gls{radiance}, ou \gls{luminance} énergétique.} 
		{is a vector function that describes the amount of light flowing in every direction through every point in space. The space of all possible light rays is given by the five-dimensional plenoptic function, and the magnitude of each ray is given by the \gls{radiance}.}
	}%
  
\newdualentry{DoF} %
	{DoF}            %
	{\iflanguage{french}{profondeur de champ}{depth of field}}  %
	{\iflanguage{french}
		{(en anglais, \textit{depth of field}), est, pour une caméra, la zone de distance dans l’espace de l'objet sur laquelle l’image d’un point objet est vue nette, \ie que sa taille sur le détecteur est inférieure à la taille du pixel du détecteur. Hors de cette zone, l'image de l'objet parait floue.} 
		{is the distance about the \acrfull{PoF} where objects appear acceptably sharp in an image.}
	}%
	
\newdualentry{CoC} %
	{CoC}            %
	{\iflanguage{french}{cercle de confusion}{circle of confusion}}  %
	{\iflanguage{french}
		{est le plus gros disque lumineux circulaire qui puisse se former sur la surface photosensible et qui sera néanmoins perçu comme un point sur le tirage final. Caractérisé par son diamètre, il permet de déterminer la limite entre ce qui est perçu flou (comme un ensemble de taches) et ce qui est perçu net (comme un ensemble de points). } 
		{is an optical spot caused by a cone of light rays from a lens not coming to a perfect focus when imaging a point source. It is also known as disk of confusion, circle of indistinctness, blur circle, or blur spot.}
	}%
	
\newdualentry{TCP}
	{TCP}            %
	{\iflanguage{french}{plan de couverture total}{total covering plane}}  %
	{\iflanguage{french}
		{toto. } 
		{is the plane, on which the main lens must be focused. If the projection of the main lens is any closer to the sensor, no depth estimation is possible.}
	}%

\newglossaryentry{mapping} %
{
	name={\iflanguage{french}{cartographie}{mapping}},  %
	description={\iflanguage{french}
		{(en anglais, \textit{mapping}), est l'action de construire ou d'améliorer une carte de son environnement. Voir \acrshort{SLAM}.}
		{is, for an agent, the ability to construct or improve a map of its environment.}
	}%
}  
\newglossaryentry{tracking} %
{
	name={\iflanguage{french}{localisation et suivi}{tracking}},  %
	description={\iflanguage{french}
	 	{(en anglais, \textit{tracking}), est l'action de suivre un ensemble d'éléments dans une séquence d'image ce qui permet de se localiser par la suite. Voir \acrshort{SLAM}. } 
	 	{describes several different methods of extracting camera motion information from a motion picture. }
	}%
} 

\newglossaryentry{aperture} %
{
	name={\iflanguage{french}{ouverture}{aperture}},  %
	description={\iflanguage{french}
		{est, pour un objectif photographique, le réglage qui permet d'ajuster le diamètre d'ouverture du diaphragme. Elle est caractérisée par le nombre d'ouverture ou ouverture géométrique, plus fréquemment notée $f/N$. À focale constante, l'augmentation du nombre d'ouverture est la conséquence de la fermeture du diaphragme : elle a pour effets la réduction de l'éclairement du capteur ou de la pellicule, l'augmentation de la profondeur de champ et, dans une moindre mesure, la réduction des aberrations géométriques et chromatiques, l'augmentation de l'influence de la diffraction.} 
		{is a hole or an opening through which light travels. More specifically, the aperture and the \gls{focal} of an optical system determine the cone angle of a bundle of rays that come to a focus in the image plane.}
	}%
}  

\newglossaryentry{focal} %
{
	name={\iflanguage{french}{focale}{focal length}},  %
	description={\iflanguage{french}
		{ou distance/longueur focale, est la distance (généralement exprimée en millimètres) qui sépare le centre optique de l’objectif et le foyer. Il s'agit d'une des caractéristiques principale d'un objectif photographique. Plus la focale sera petite plus le \acrlong{FOV} sera grand.} 
		{is, for an optical system, a measure of how strongly the system converges or diverges light.}
	}%
}  

\newglossaryentry{magnification} %
{
	name={\iflanguage{french}{grandissement}{magnification}},  %
	description={\iflanguage{french}
		{(en anglais, \textit{magnification}) est associé au rapport d'une grandeur de l'objet à son équivalent pour l'image de cet objet à travers un système optique. On distingue les grandissements transversal, angulaire, longitudinal et pupillaire.} 
		{is the process of enlarging the apparent size, not physical size, of something. This enlargement is quantified by a calculated number also called magnification. When this number is less than one, it refers to a reduction in size, sometimes called \textit{minification} or \textit{de-magnification}.}
	}%
}  

\newglossaryentry{luminance} %
{
	name={\iflanguage{french}{luminance}{luminance}},  %
	description={\iflanguage{french}
		{est une grandeur correspondant à la sensation visuelle de luminosité d'une surface. Une surface très lumineuse présente une forte luminance, tandis qu'une surface parfaitement noire aurait une luminance nulle.} 
		{is a photometric measure of the luminous intensity per unit area of light traveling in a given direction. It describes the amount of light that passes through, is emitted or reflected from a particular area, and falls within a given solid angle. }
	}%
}  

\newglossaryentry{radiance} %
{
	name={\iflanguage{french}{radiance}{radiance}},  %
	description={\iflanguage{french}
		{aussi \gls{luminance} énergétique, est la puissance par unité de surface du rayonnement passant ou étant émis en un point d'une surface, et dans une direction donnée par unité d'angle solide.} 
		{is the radiant flux emitted, reflected, transmitted or received by a given surface, per unit solid angle per unit projected area.}
	}%
}  

\newglossaryentry{vignetting} %
{
	name={\iflanguage{french}{vignettage}{vignetting}},  %
	description={\iflanguage{french}
		{(en anglais, \textit{vignetting}), est la diminution de l'intensité et/ou de la saturation à la périphérie d'une image par rapport à son centre donnant l'impression de coins sombres.} 
		{is a reduction of an image's brightness or saturation toward the periphery compared to the image center.}
	}%
}

\setabbreviationstyle[acronym]{long-short}

\newacronym{VO}{VO}{\iflanguage{french}{odométrie visuelle}{Visual Odometry}}
\newacronym{VIO}{VIO}{\iflanguage{french}{odométrie visuelle-inertielle}{Visual-Inertial Odometry}}

\newacronym{LIDAR}{LIDAR}{\iflanguage{french}{mesure de distance laser}{Light Detection and Ranging}}
\newacronym{TOF}{ToF}{\iflanguage{french}{temps de vol}{Time of Flight}}

\newacronym{PSF}{PSF}{\iflanguage{french}{fonction d'étalement du point}{Point-Spread Function}}
\newacronym{OTF}{OTF}{\iflanguage{french}{fonction de tranfert optique}{Optical Transfer Function}}
\newacronym{MTF}{MTF}{\iflanguage{french}{fonction de tranfert de modulation}{Modulation Transfer Function}}
\newacronym{PhTF}{PhTF}{\iflanguage{french}{fonction de tranfert de phase}{Phase Transfer Function}}
\newacronym{CTF}{CTF}{\iflanguage{french}{fonction de tranfert de constraste}{Contrast Transfer Function}}
\newacronym{LSF}{LSF}{\iflanguage{french}{fonction d'étalement de ligne}{Line-Spread Function}}

\newacronym{PoF}{PoF}{\iflanguage{french}{plan de focalisation}{Plane of Focus}}
\newacronym{LFC}{LFC}{\iflanguage{french}{caméra plénoptique}{Light Field Camera}}
\newacronym{SPC}{SPC}{\iflanguage{french}{caméra plenoptique standard}{Standard Plenoptic Camera}}
\newacronym{MI}{MI}{\iflanguage{french}{micro-image}{micro-image}}
\newacronym{MIC}{MIC}{\iflanguage{french}{centre de la micro-image}{micro-image center}}
\newacronym{MLA}{MLA}{\iflanguage{french}{matrice de micro-lentilles}{Micro-Lenses Array}}
\newacronym{EPI}{EPI}{\iflanguage{french}{image des plans épipolaires}{Epipolar-Planes Image}}
\newacronym{MIA}{MIA}{\iflanguage{french}{matrice de micro-images}{Micro-Images Array}}
\newacronym{SAI}{SAI}{\iflanguage{french}{image à ouverture partielle}{sub-aperture image}}
\newacronym{ULFC}{ULFC}{\iflanguage{french}{caméra plénoptique non-focalisée}{Unfocused Light Field Camera}}
\newacronym{FLFC}{FLFC}{\iflanguage{french}{caméra plénoptique focalisée}{Focused Light Field Camera}}
\newacronym{MFPC}{MFPC}{\iflanguage{french}{caméra plénoptique multi-focale}{Multi-Focus Plenoptic Camera}}
\newacronym{BAP}{BAP}{\iflanguage{french}{tennant compte du flou plénoptique}{Blur Aware Plenoptic}}

\newacronym{RMSE}{RMSE}{\iflanguage{french}{Écart quadratique moyen}{Root-Mean-Square Error}}

\newacronym{FPS}{fps}{\iflanguage{french}{image par seconde}{frame per second}}

\newacronym{GPU}{GPU}{\iflanguage{french}{processeur graphique}{Graphics Processing Unit}}
\newacronym{FPGA}{FPGA}{\iflanguage{french}{circuit logique programmable}{Field-Programmable Gate Array}}
\newacronym{API}{API}{\iflanguage{french}{interface de programmation applicative}{Application Programming Interface}}

\newacronym{PIV}{PIV}{\iflanguage{french}{vélocimétrie par images de particules}{Particle Image Velocimetry}}

\newacronym{GAN}{GANs}{\iflanguage{french}{réseaux adverses génératifs}{Generative Adversarial Networks}}

\newacronym{SfM}{SfM}{structure from motion}
\newacronym{RANSAC}{RANSAC}{Random Sample Concensus}
\newacronym{PNP}{PnP}{Perspective-n-Point}
\newacronym{DBSCAN}{DBSCAN}{Density-Based Spatial Clustering of Applications with Noise}

\newacronym{PCA}{PCA}{\iflanguage{french}{Analyse en Composantes Principales}{Principal Component Analysis}}
\newacronym{SVD}{SVD}{\iflanguage{french}{décomposition en valeurs singulières}{Singular-Value Decomposition}}

%% file: figures/tex/notations-all.tex
\begin{tikzpicture}[scale=\tkzscl]
    \tikzset{
        lens/.pic={ \draw [xscale=.2] (0,1) .. controls (1,.5) and (1,-.5) .. (0,-1) .. controls (-1,-.5) and (-1,.5) .. cycle; },
        lightlens/.pic={ \draw [gray,dashed,xscale=.2] (0,1) .. controls (1,.5) and (1,-.5) .. (0,-1) .. controls (-1,-.5) and (-1,.5) .. cycle; },
    }
	\tikzstyle{point} = [circle, fill, inner sep = 1pt]
	
    \newcommand*\frametikz[4]
    {
        \node[above, yshift = 1.0em] at (#2, #3) {};
        \draw[] (#2, #3) circle (5pt) node[above left, xshift = 0.2em]{$x\fworld$};
        \filldraw (#2, #3) node[point] {};
        \draw[line width = 0.5pt, color = black, ->] (#2, #3) --
        (#2, #3 - #4) node[left, xshift = 0.2em] {$y\fworld$};
        \draw[line width = 0.5pt, color = black, ->] (#2, #3) --
        (#2 + #4, #3) node[above] {$z\fworld$};
    };%
    \draw[dashed] (-0.1,0) -- (10.1,0); 

    \draw (1.5,-3) -- (1.5,3);
    \node at (1.5,3.2) {Sensor};

    \draw[thick,violet,->] (0,0) -- (0,-0.8) node[below] {image};
    \draw[<->] (0,-0.8) --node[above,pos = 0.8]{$\distimg\flens$} (7,-0.8);
    \draw[thick,violet,->] (10,0) -- (10,1.5) node [above] {object};
    \draw[<-] (7,-0.8) -- (8.45,-0.8) node{/};
    \draw[->] (8.55,-0.8) node{/} --node[above]{$\distobj\flens$} (10,-0.8);

    \path (7,0) pic [scale=\tkzscl*2] {lens};
    \node at (7.25,2.2) {Main Lens};
    \draw[<->] (7.5,-2) --node[right,pos = 0.75]{$\aperture$} (7.5,2);
    \draw[thick] (7,0.1) --node[above]{$\lenscenter$}(7,-0.1);

    \path (3,0) foreach \dy in {0,...,2} {+(0,\dy) pic [scale=\tkzscl*0.5] {lens}};
    \path (3,0) foreach \dy in {-2,...,-1} {+(0,\dy) pic [scale=\tkzscl*0.5] {lightlens}};
    \node at (3,2.7) {\gls{MLA}};
    
    \draw[red] (7,0) -- (1.5,2.666);
    \draw[red] (7,0) -- (1.5,1.333);
    \draw[fill=black] (1.5,2.666) circle (1pt);
    \draw[fill=black] (1.5,1.333) circle (1pt);

    \draw[thick, green] (1,0.5) -- (3.5,0.5) -- (3.5,1.5) -- (1,1.5) -- cycle;
    
    \draw[fill=black] (1.5,0) circle (2pt) node[below]{$\Principalpoint$};
    
    \draw[thick] (0.4,-0.1) -- (0.6,0.1);
    \draw[thick](0.4,0.1) -- (0.6,-0.1);
    
    \draw[<->] (1.5,-1.4) --node[above]{$\distmlasensor$} (3,-1.4);
    \draw[<->] (3,-1.4) --node[above]{$\distlensmla$} (7,-1.4);
    \draw[<->] (0.5,-2) --node[above]{$\Focal$} (7,-2);
    \draw[<->] (1.25,1.33) --node[left]{$\miinterdistmm$} (1.25,2.66);
    
    \frametikz{}{9}{3}{0.5};

    \draw[green] (4.5,-2.125) -- (9.5,-2.125) -- (9.5,-4.5) -- (4.5,-4.5) -- cycle;
    \draw[dashed,green] (3.5,0.5) -- (4.5,-2.125);
    \node at (7,-3.5) {\input{figures/tex/notations-mla.tex}};
    
\end{tikzpicture}

%% file: figures/tex/miradius.tex
\begin{tikzpicture}[scale=\tkzscl, every node/.style={scale=0.9}] %
	\tikzset{
		basefont/.style = {font = \small},
		timing/.style = {basefont, sloped,above,},
		label/.style = {basefont, align = left},
		lens/.pic={ \draw [xscale=.2] (0,1) .. controls (1,.5) and (1,-.5) .. (0,-1) .. controls (-1,-.5) and (-1,.5) .. cycle; },
	};
	
	\coordinate (bot_sensor) at (0,-3.1313559322033893);
	\coordinate (top_sensor) at (0,2.1313559322033893);
	\coordinate (bot_R) at ($ (bot_sensor) + (-0.25,0) $);
	\coordinate (top_R) at ($ (top_sensor) + (-0.25,0) $);
	\coordinate (mid_R) at (-0.25,-0.488647094);
	\coordinate (bot_sensor_plan) at (0,-4);
	\coordinate (top_sensor_plan) at (0,4);
	
	\coordinate (bot_MLA) at (5,-1);
	\coordinate (top_MLA) at (5,2);
	\coordinate (bot_MLA_size) at ($ (bot_MLA) + (0.35,0) $);
	\coordinate (top_MLA_size) at ($ (top_MLA) + (0.35,0) $);
	\coordinate (bot_MLA_plan) at (5,-4);
	\coordinate (top_MLA_plan) at (5,4);
	
	\coordinate (bot_ML) at (10,-1.125);
	\coordinate (mid_ML) at (10,1.5);
	\coordinate (top_ML) at (10,4.125);
	\coordinate (bot_ML_size) at ($ (bot_ML) + (0.25,0) $);
	\coordinate (top_ML_size) at ($ (top_ML) + (0.25,0) $);
	
	\coordinate (V) at (-1.6666666666666667,-0.8333333333333334);
	\coordinate (V') at (3.1846153846153844,0.13692307692307687);
	\coordinate (foc) at (2,0.5);
	\coordinate (left_foc) at (0,0.5);
	\coordinate (right_foc) at (5,0.5);
	
	\coordinate (left_b') at (3.1846153846153844,-1.5);
	\coordinate (right_b') at (5,-1.5);
	
	\coordinate (left_f) at (2,-2.3);
	\coordinate (right_f) at (5,-2.3);
	
	\coordinate (left_d) at (0,-3);
	\coordinate (right_d) at (5,-3);
	
	\coordinate (left_D) at (5,-3);
	\coordinate (right_D) at (10,-3);
	
	\coordinate (left_a') at (-1.6666666666666667,-3.7);
	\coordinate (right_a') at (5,-3.7);

	\draw[dashed] (mid_ML) -- (V);
	\draw[gray,dashed] (left_foc) -- (right_foc);
	
	\draw[very thick,green] (bot_sensor)-- (top_MLA);
	\draw[green] (V)-- (top_ML);
	
	\draw[red] (V)-- (bot_ML);
	\draw[very thick,red] (top_sensor)-- (bot_MLA);
	
	\draw[fill=black] (V) circle (2pt) node[below]{$V$};
	\draw[fill=black] (V') circle (2pt);%
	\node at ($(V') + (+0.1, -0.50)$) {$V'$};
	\draw[thick] ($(foc) + (-0.1,-0.1)$) -- ($(foc) + (0.1,0.1)$);
	\draw[thick] ($(foc) + (-0.1,0.1)$) -- ($(foc) + (0.1,-0.1)$);
	
	\draw[] (top_ML) --node[below,pos=1]{Main Lens} (bot_ML);
	\draw[<->, thick] (top_ML) --node{-} (bot_ML);
	\draw[<->] (top_ML_size)--node[right]{$\aperture$} (bot_ML_size);
	
	\draw[dashed] (top_MLA_plan) --node[left,pos=0.1]{\gls{MLA}} (bot_MLA_plan);
	\path ($(top_MLA)!0.5!(bot_MLA)$) pic [scale=1] {lens};
	\draw[<->] (top_MLA_size) --node[right,pos=0.3]{$\mlinterdist$} (bot_MLA_size);
	
	\draw[dashed] (top_sensor_plan) --node[right,pos=0.1]{Sensor} (bot_sensor_plan);
	\draw[very thick] (top_sensor) --node{-} (bot_sensor);
	\draw[<->] (top_R) --node[left]{$\miradiusmm$} (mid_R);
	\draw[<->] (bot_R) --node[left]{$\miradiusmm$} (mid_R);

	\draw[<->] (left_b') --node[below]{$\distimg'$} (right_b');
	\draw[<->] (left_f) --node[below]{$\focal$} (right_f);
	\draw[<->] (left_d) --node[below]{$\distmlasensor$} (right_d);
	\draw[<->] (left_D) --node[below]{$\distlensmla$} (right_D);
	\draw[<->] (left_a') --node[below]{$\distobj'$} (right_a');
\end{tikzpicture}

%% file: root.bbl
\begin{thebibliography}{10}\itemsep=-1pt

\bibitem{Adelson1991}
E.~H. Adelson and J.~R. Bergen.
\newblock {The plenoptic function and the elements of early vision}.
\newblock {\em Computational Models of Visual Processing}, pages 3--20, 1991.

\bibitem{Bok2014}
Yunsu Bok, Hae-Gon Jeon, and In~So Kweon.
\newblock {Geometric Calibration of Micro-Lens-Based Light-Field Cameras Using
  Line Features}.
\newblock In {\em Computer Vision -- ECCV 2014}, pages 47--61. Springer
  International Publishing, 2014.

\bibitem{Bok2017}
Yunsu Bok, Hae-Gon Jeon, and In~So Kweon.
\newblock {Geometric Calibration of Micro-Lens-Based Light Field Cameras Using
  Line Features}.
\newblock {\em IEEE Transactions on Pattern Analysis and Machine Intelligence},
  39(2):287--300, 2017.

\bibitem{Brown1966}
Duane Brown.
\newblock {Decentering Distortion of Lenses - The Prism Effect Encountered in
  Metric Cameras can be Overcome Through Analytic Calibration}.
\newblock {\em Photometric Engineering}, 32(3):444--462, 1966.

\bibitem{Conrady1919}
Ae Conrady.
\newblock {Decentered Lens-Systems}.
\newblock {\em Monthly Notices of the Royal Astronomical Soceity}, 79:384--390,
  1919.

\bibitem{Dansereau2013c}
Donald~G. Dansereau, Oscar Pizarro, and Stefan~B. Williams.
\newblock {Decoding, calibration and rectification for lenselet-based plenoptic
  cameras}.
\newblock {\em Proceedings of the IEEE Computer Society Conference on Computer
  Vision and Pattern Recognition}, pages 1027--1034, 2013.

\bibitem{Ester1996}
Martin Ester, Hans-Peter Kriegel, Jiirg Sander, and Xiaowei Xu.
\newblock {A} {Density-Based Algorithm for Discovering Clusters in Large
  Spatial Databases with Noise}.
\newblock In {\em KDD}, 1996.

\bibitem{Georgiev2009e}
Todor Georgiev and Andrew Lumsdaine.
\newblock {Resolution} in {Plenoptic Cameras}.
\newblock {\em Frontiers in Optics 2009/Laser Science XXV/Fall 2009 OSA Optics
  \& Photonics Technical Digest}, page CTuB3, 2009.

\bibitem{Georgiev2012}
Todor Georgiev and Andrew Lumsdaine.
\newblock {The multifocus plenoptic camera}.
\newblock In {\em Digital Photography VIII}, pages 69--79. International
  Society for Optics and Photonics, SPIE, 2012.

\bibitem{Georgiev2006}
Todor Georgiev, Ke~Colin Zheng, Brian Curless, David Salesin, Shree~K Nayar,
  and Chintan Intwala.
\newblock {Spatio-Angular Resolution Tradeoff in Integral Photography}.
\newblock {\em Rendering Techniques}, 2006(263-272):21, 2006.

\bibitem{Hahne2018}
Christopher Hahne, Andrew Lumsdaine, Amar Aggoun, and Vladan Velisavljevic.
\newblock {Real-time refocusing using an FPGA-Based standard plenoptic camera}.
\newblock {\em IEEE Transactions on Industrial Electronics}, 65(12):9757--9766,
  2018.

\bibitem{Heinze2014}
Christian Heinze.
\newblock {\em {Design and test of a calibration method for the calculation of
  metrical range values for 3D light field cameras}}.
\newblock Master's thesis, Hamburg University of Applied Sciences - Faculty of
  Engineering and Computer Science, 2014.

\bibitem{Heinze2016b}
Christian Heinze, Stefano Spyropoulos, Stephan Hussmann, and Christian
  Perwa{\ss}.
\newblock {Automated Robust Metric Calibration Algorithm for Multifocus
  Plenoptic Cameras}.
\newblock {\em IEEE Transactions on Instrumentation and Measurement},
  65(5):1197--1205, 2016.

\bibitem{Johannsen2013b}
Ole Johannsen, Christian Heinze, Bastian Goldluecke, and Christian Perwa{\ss}.
\newblock {On the calibration of focused plenoptic cameras}.
\newblock {\em Lecture Notes in Computer Science (including subseries Lecture
  Notes in Artificial Intelligence and Lecture Notes in Bioinformatics)}, 8200
  LNCS:302--317, 2013.

\bibitem{Kneip2011b}
Laurent Kneip, Davide Scaramuzza, and Roland Siegwart.
\newblock {A novel parametrization of the perspective-three-point problem for a
  direct computation of absolute camera position and orientation}.
\newblock {\em Proceedings of the IEEE Computer Society Conference on Computer
  Vision and Pattern Recognition}, pages 2969--2976, 2011.

\bibitem{Levin2008b}
Anat Levin, William~T. Freeman, and Fr{\'{e}}do Durand.
\newblock {Understanding camera trade-offs through a Bayesian analysis of light
  field projections}.
\newblock {\em Lecture Notes in Computer Science (including subseries Lecture
  Notes in Artificial Intelligence and Lecture Notes in Bioinformatics)}, 5305
  LNCS(PART 4):88--101, 2008.

\bibitem{Lippmann1911b}
Gabriel Lippmann.
\newblock {Integral Photography}.
\newblock {\em Academy of the Sciences}, 1911.

\bibitem{Lumsdaine2009b}
Andrew Lumsdaine and Todor Georgiev.
\newblock {The focused plenoptic camera}.
\newblock In {\em IEEE International Conference on Computational Photography
  (ICCP)}, pages 1--8, apr 2009.

\bibitem{Ng2005b}
Ren Ng, Marc Levoy, Gene Duval, Mark Horowitz, and Pat Hanrahan.
\newblock {Light Field Photography with a Hand-held Plenoptic Camera}.
\newblock Technical report, Stanford University, 2005.

\bibitem{Noury2017b}
Charles~Antoine Noury, C{\'{e}}line Teuli{\`{e}}re, and Michel Dhome.
\newblock {Light-Field Camera Calibration from Raw Images}.
\newblock {\em DICTA 2017 -- International Conference on Digital Image
  Computing: Techniques and Applications}, pages 1--8, 2017.

\bibitem{Nousias2017}
Sotiris Nousias, Francois Chadebecq, Jonas Pichat, Pearse Keane, Sebastien
  Ourselin, and Christos Bergeles.
\newblock {Corner-Based Geometric Calibration of Multi-focus Plenoptic
  Cameras}.
\newblock {\em Proceedings of the IEEE International Conference on Computer
  Vision}, pages 957--965, 2017.

\bibitem{OBrien2018}
Sean O'Brien, Jochen Trumpf, Viorela Ila, and Robert Mahony.
\newblock {Calibrating light-field cameras using plenoptic disc features}.
\newblock In {\em 2018 International Conference on 3D Vision (3DV)}, pages
  286--294. IEEE, 2018.

\bibitem{Perwass2010c}
Christian Perwa{\ss} and Lennart Wietzke.
\newblock {Single Lens 3D-Camera with Extended Depth-of-Field}.
\newblock In {\em Human Vision and Electronic Imaging XVII}, volume~49, page
  829108. SPIE, feb 2012.

\bibitem{Shi2019}
Shengxian Shi, Junfei Ding, T.~H. New, You Liu, and Hanmo Zhang.
\newblock {Volumetric calibration enhancements for single-camera light-field
  PIV}.
\newblock {\em Experiments in Fluids}, 60(1):21, jan 2019.

\bibitem{Shi2016}
Shengxian Shi, Jianhua Wang, Junfei Ding, Zhou Zhao, and T.~H. New.
\newblock {Parametric study on light field volumetric particle image
  velocimetry}.
\newblock {\em Flow Measurement and Instrumentation}, 49:70--88, 2016.

\bibitem{Strobl2016}
Klaus~H. Strobl and Martin Lingenauber.
\newblock {Stepwise calibration of focused plenoptic cameras}.
\newblock {\em Computer Vision and Image Understanding}, 145:140--147, 2016.

\bibitem{Subbarao1994}
Murali Subbarao and Gopal Surya.
\newblock {Depth from defocus: A spatial domain approach}.
\newblock {\em International Journal of Computer Vision}, 13(3):271--294, 1994.

\bibitem{Suliga2018}
Piotr Suliga and Tomasz Wrona.
\newblock {Microlens array calibration method for a light field camera}.
\newblock {\em Proceedings of the 19th International Carpathian Control
  Conference (ICCC)}, pages 19--22, 2018.

\bibitem{Sun2016}
Jun Sun, Chuanlong Xu, Biao Zhang, Shimin Wang, Md~Moinul Hossain, Hong Qi, and
  Heping Tan.
\newblock Geometric calibration of focused light field camera for {3-D} flame
  temperature measurement.
\newblock In {\em Conference Record - IEEE Instrumentation and Measurement
  Technology Conference}, July 2016.

\bibitem{Thomason2014}
Chelsea~M. Thomason, Brian~S. Thurow, and Timothy~W. Fahringer.
\newblock {Calibration of a Microlens Array for a Plenoptic Camera}.
\newblock {\em 52nd Aerospace Sciences Meeting}, (January):1--18, 2014.

\bibitem{Wang2018}
Yuan Wang, Jun Qiu, Chang Liu, Di He, Xinkai Kang, Jian Li, and Ligen Shi.
\newblock {Virtual Image Points Based Geometrical Parameters' Calibration for
  Focused Light Field Camera}.
\newblock {\em IEEE Access}, 6(c):71317--71326, 2018.

\bibitem{Zeller2016h}
Niclas Zeller, Charles~Antoine Noury, Franz Quint, C{\'{e}}line Teuli{\`{e}}re,
  Uwe Stilla, and Michel Dhome.
\newblock {Metric Calibration of a Focused Plenoptic Camera based on a 3D
  Calibration Target}.
\newblock {\em ISPRS Annals of Photogrammetry, Remote Sensing and Spatial
  Information Sciences}, III-3(July):449--456, jun 2016.

\bibitem{Zeller2014}
Niclas Zeller, Franz Quint, and Uwe Stilla.
\newblock {Calibration and accuracy analysis of a focused plenoptic camera}.
\newblock {\em ISPRS Annals of Photogrammetry, Remote Sensing and Spatial
  Information Sciences}, II-3(September):205--212, 2014.

\bibitem{Zhang2016}
Chunping Zhang, Zhe Ji, and Qing Wang.
\newblock {Decoding and calibration method on focused plenoptic camera}.
\newblock {\em Computational Visual Media}, 2(1):57--69, 2016.

\bibitem{Zhang2016a}
Chunping Zhang, Zhe Ji, and Qing Wang.
\newblock {Unconstrained Two-parallel-plane Model for Focused Plenoptic Cameras
  Calibration}.
\newblock pages 1--20, 2016.

\bibitem{Zhou2019}
Ping Zhou, Weijia Cai, Yunlei Yu, Yuting Zhang, and Guangquan Zhou.
\newblock {A two-step calibration method of lenslet-based light field cameras}.
\newblock {\em Optics and Lasers in Engineering}, 115:190--196, 2019.

\end{thebibliography}
